\documentclass[%
 reprint,
superscriptaddress,
amsmath,amssymb,
aps,
prb,
longbibliography
]{revtex4-1}
\usepackage{graphicx}% Include figure files
\usepackage{dcolumn}% Align table columns on decimal point
\usepackage{bm}% bold math
\usepackage{multirow}
\usepackage{amsmath}%
\allowdisplaybreaks
\usepackage{bbm}
\usepackage{notes2bib}
\usepackage{float}
\usepackage{xcolor}
\usepackage[colorlinks=true,linkcolor=blue,citecolor=blue]{hyperref}

\newcommand{\bd}[1]{\mathbf{#1}}
\newcommand{\bra}[1]{\left\langle#1\right|}
\newcommand{\ket}[1]{\left|#1\right\rangle}

\begin{document}

\preprint{APS/123-QED}

\title{Fractional disclination charge in two-dimensional $C_n-$symmetric topological crystalline insulators}% Force line breaks with \\
%\thanks{A footnote to the article title}%

\author{Tianhe Li}
\author{Penghao Zhu}\altaffiliation{Tianhe Li and Penghao Zhu contributed equally to this work.}
\affiliation{%
 Department of Physics, Institute for Condensed Matter Theory,\\
University of Illinois at Urbana-Champaign, IL 61801, USA
}

\author{Wladimir A. Benalcazar}
\affiliation{Department of Physics, The Pennsylvania State University, University Park, PA 16801, USA}
\author{Taylor L. Hughes}
% \email{Second.Author@institution.edu}
\affiliation{%
 Department of Physics, Institute for Condensed Matter Theory,\\
University of Illinois at Urbana-Champaign, IL 61801, USA
}

\date{\today}% It is always \today, today,
             %  but any date may be explicitly specified

\begin{abstract}
Robust fractional charge localized at disclination defects has been recently found as a topological response in $C_{6}$ symmetric 2D topological crystalline insulators (TCIs). 
In this article, we thoroughly investigate the fractional charge on disclinations in $C_n$ symmetric TCIs, with or without time reversal symmetry, and including spinless and spin-$\frac{1}{2}$ cases. 
We compute the fractional disclination charges from the  Wannier representations in real space and use band representation theory to construct topological indices of the fractional disclination charge for all $2D$ TCIs that admit a (generalized) Wannier representation.
We find the disclination charge is fractionalized in units of $\frac{e}{n}$ for $C_n$ symmetric TCIs; and for spin-$\frac{1}{2}$ TCIs, with additional time reversal symmetry, the disclination charge is fractionalized in units of $\frac{2e}{n}$.
We furthermore prove that with electron-electron interactions that preserve the $C_n$ symmetry and many-body bulk gap, though we can deform a TCI into another which is topologically distinct in the free fermion case, the fractional disclination charge determined by our topological indices will not change in this process.
Moreover, we use an algebraic technique to generalize the indices for TCIs with non-zero Chern numbers, where a Wannier representation is not applicable. With the inclusion of the Chern number, our generalized fractional disclination indices apply for all $C_n$ symmetric TCIs. 
Finally, we briefly discuss the connection between the Chern number dependence of our generalized indices and the Wen-Zee term.
\end{abstract}

\maketitle

\section{Introduction}
Topological insulators (TIs) are insulators having a gapped bulk, but metallic boundaries~\cite{hasan2010,RevModPhys.83.1057,bernevig2013book,Ludwig_2015}. 
Strong TIs with local symmetries (time reversal symmetry, particle-hole symmetry and/or chiral symmetry) host gapless states at the boundaries of co-dimension $1$, e.g., edges in 2D and surfaces in 3D. 
The TI phases have quantized topological invariants defined in the bulk and are robust to perturbations, including disorder, that preserve the symmetry and the bulk energy gap~\cite{RevModPhys.83.1057}.
In the past decade, studies of TIs have been extended to insulators with additional crystalline symmetries, such as mirror or inversion symmetries~\cite{fu2007topological,teo2008surface,turner2010,fu2011topological,hughes2011inversion,hsieh2012,Yoichi2015Topological,Ching-Kai2016Classification}. 
Conventionally, TIs with gapless boundary states that are protected by crystalline symmetries are referred to as topological crystalline insulators (TCIs).
Recently, a new class of TCIs has been discovered -- they are gapped at boundaries with codimension $1,$ but have robust, symmetry-protected features at boundaries with higher codimension, e.g., corners in $2D$~\cite{benalcazar2017quantized,Langbehn2017Reflection}, and hinges or corners in $3D$~\cite{Langbehn2017Reflection,benalcazar2018prb,Schindlereaat0346,schindler2018higher}. Such insulators are known as higher-order topological insulators (HOTIs).
In $2D$ TCIs with gapped edges, crystalline symmetries alone \emph{cannot} pin the corner states to the middle of a band gap, so they are not a generic topological feature\cite{benalcazar2019quantization,ahn2019failure}. In order to have protected mid-gap states, additional particle-hole or chiral symmetries are necessary~\cite{benalcazar2019quantization,ahn2019failure}. As such, a generic topological character of  TCIs without protected, mid-gap boundary modes is the quantization of fractional charge at boundaries~\cite{SSHmodel,fang2012bulk,van2018higher,benalcazar2017quantized,benalcazar2019quantization,wieder2018axion,lee2019fractional,lee2019higher,PhysRevResearch.1.033074} or odd-parity integer charge at boundaries in spin-$\frac{1}{2}$ systems with time reversal symmetry\cite{song2017d,Alexander2016Topological}. In the bulk, non-trivial topology due to crystalline symmetries can typically be diagnosed by the symmetry representations of the occupied bands at the high symmetry points (HSP) in the momentum space Brillouin zone (BZ)~\cite{fu2007topological,fang2012bulk,bradlyn2017topological,po2017symmetry,JenniferBuilding2018,bradlyn2019disconnected,kruthoff2017topological}. These representations correspond to (generalized) configurations of Wannier orbitals (defined in Sec.~\ref{Sec: fracdisWan}) in real space through band representation theory\cite{bradlyn2017topological,po2017symmetry,JenniferBuilding2018,bradlyn2019disconnected,kruthoff2017topological}.
It has been found that in TCIs with rotation symmetries, a topologically non-trivial (generalized) configuration of Wannier orbitals may lead to quantized fractional charges that are exponentially localized at the corners\cite{benalcazar2017quantized,van2018higher,benalcazar2019quantization,wieder2018axion,lee2019fractional,lee2019higher,PhysRevResearch.1.033074}. 

In this article we explore the existence of robust fractional charge on disclination defects in crystals with rotation symmetries\footnote{We note that we exclude the exceptional cases in which integer topological charge is present because of the lack of symmetry indicators to diagnose them.}. These defects have been shown to give rise to topological bound states in topological crystalline superconductors protected by rotation symmetries\cite{PhysRevLett.111.047006,benalcazar2014classification}, so it is natural to also try to identify any manifestations of fractional charge in TCIs.
To motivate the idea, consider, for example, a $C_4$ symmetric TCI on an open square lattice having a fractional corner charge, where the fractional charges from all four corners add up to an integer. One can form a disclination through a cutting and gluing procedure. First, remove one quadrant of the lattice, and then glue the cut edges together. This will generate a disclination at the center of the lattice. From this construction there are now only three corners in the lattice so the sum of charges at the remaining corners is fractional. This implies that there will be a fractional charge at the core of the disclination since the total charge in the lattice must take an integer value. 

Motivated by this intuition, we expect that disclinations can generally bind quantized fractional charges.  
Indeed, recent studies have proposed models of $C_6$ symmetric TCIs that exhibit fractional charge localized on disclination cores\cite{benalcazar2019quantization,liu2018shift}.
Additionally, in a $C_{6}$ symmetric Chern insulator (which does not have a Wannier orbital description),
numerical calculations reveal the existence of quantized fractional charges localized at the disclination cores~\cite{liu2018shift,he2018chern}.
Despite these recent findings, a general topological index for the existence of quantized fractional disclination charges in $C_n$ symmetric $2D$ TCIs is still missing.

In this article, we initially consider $2D$ $C_n$-symmetric TCIs that have (generalized) Wannier representations and derive the origin of charge fractionalization at disclination cores through a real space Wannier orbital picture (Sec.~\ref{Sec: fracdisWan}).
From this information, and using the theory of band representations~\cite{bradlyn2017topological,po2017symmetry,JenniferBuilding2018,bradlyn2019disconnected,kruthoff2017topological}, we build the indices of fractional disclination charge for all TCIs with zero Chern number (Sec.~\ref{sec:fracdisrotidx}). 
These indices apply to all $2D$ TCIs that admit a (generalized) Wannier orbital description, with or without TRS, but they are not capable of capturing the fractional disclination charge in insulators with non-zero Chern number.
Then we allow for symmetry-preserving electron-electron interactions and find that the fractional charge determined by the non-interacting TCI indices remains robust as long as the interacting phases are adiabatically connected to non-interacting TCIs (Sec. \ref{sec:interaction}).
Finally, in Sec.~\ref{Sec:formulawithch}, we extend the indices constructed in Sec.~\ref{sec:fracdisrotidx} using an algebraic technique\cite{benalcazar2014classification,PhysRevLett.111.047006} to build the generalized indices for all 2D $C_n$ symmetric TCIs, including those with Chern numbers, and both spinless and spin-$\frac{1}{2}$ cases. 

In general, we find that the charge is fractionalized at disclination cores in units of $\frac{e}{n}$ for the $C_n$ symmetric spinless TCIs, which is in agreement with recent findings for the quantization of fractional corner charge in $C_n$ symmetric HOTIs~\cite{benalcazar2019quantization}, and in units of $\frac{2e}{n}$ for the $C_n$ symmetric spin-$\frac{1}{2}$ TCIs with additional time reversal symmetry (TRS). Although a complete one-to-one correspondence between (generalized) configurations of Wannier orbitals in real space and symmetry representations at HSPs in the ${\bf k}$ space is absent, we show that the combination of symmetry representations themselves in the BZ and the Chern number can uniquely determine the fractional disclination charges. 

\section{Disclinations in $2D$ Insulators}
\label{Sec:Disclination}
A two-dimensional $C_n$ symmetric lattice can be divided into $n$ sectors, that each subtend an angle of $\frac{2\pi}{n}$ from the rotation center, and which are related to the others by $C_n$ rotations.  Inserting or removing such sectors into/from the lattice creates a $0D$ defect of the discrete $C_n$ rotation symmetry, called a disclination. In Fig.~\ref{fig:ConstructDis}, we show examples of disclinations constructed from a $C_4$ symmetric square lattice. 
Figure~\ref{fig:ConstructDis}(a) shows the perfect lattice with $C_4$ rotation and lattice translation symmetry. The shadowed area indicates the four sectors that are related by $C_4$ rotations. Figure~\ref{fig:ConstructDis}(b,d) correspond to removing a $90^\circ$ sector of the lattice, while Fig.~\ref{fig:ConstructDis}(c) corresponds to inserting two $90^\circ$ sectors into the lattice.
A disclination is characterized by the holonomy of a closed path around its core, i.e., the amount of translation and rotation accumulated by a vector after being parallel transported along the loop. We denote the holonomy by the operator $\hat{t}_{\bf a}\hat{r}(\Omega)$, where $\hat{r}$ is the discrete rotation operator and $\Omega$ is called the Frank angle, while $\hat{t}$ is the discrete translation operator and $\bd{a}$ is a lattice vector: 
${\bf a}=a_1 {\bf e}_1+a_2{\bf e}_2$
\cite{benalcazar2014classification}.
For $C_2$ and $C_4$ symmetric lattices, we choose the unit lattice basis vectors to be  ${\bf e}_1=(1,0)$ and ${\bf e}_2=(0,1),$ and for $C_3$ and $C_6$ symmetric lattices we choose the lattice vectors to be ${\bf e}_1=(1,0)$ and ${\bf e}_2=(\frac{1}{2},\frac{\sqrt{3}}{2})$, as shown in Fig.~\ref{fig:WyckoffPos}. 
For example, in Fig.~\ref{fig:ConstructDis}(b), if we start from the open red circles and follow the direction of the arrow, after one loop, the holonomy is $\hat{r}(\frac{\pi}{2})t_{2{\bf e}_1}\hat{r}(\frac{\pi}{2})t_{2{\bf e}_1}\hat{r}(\frac{\pi}{2})t_{2{\bf e}_1}=t_{-2{\bf e}_1}\hat{r}(-\frac{\pi}{2})$.  In the presence of a $C_n$ symmetry, the compatible Frank angles, are multiples of the minimum Frank angle $\frac{2\pi}{n}$.

In general, while the Frank angle in the holonomy does not depend on the choice of the closed loop (as long as the orientation of the loop enclosing the defect is fixed), the translation part ${\bf a}$ does depend on the starting point of the closed path. Upon a translation of the starting point by $\hat{t}_{\bf c}$, the holonomy becomes
\begin{align}
\hat{t}_{\bf a}\hat{r}(\Omega)\rightarrow\hat{t}_{\bf c}\hat{t}_{\bf a}\hat{r}(\Omega)\hat{t}_{-\bf c}=\hat{t}_{{\bf a}+(1-\hat{r}(\Omega)){\bf c}}\hat{r}(\Omega),
\end{align}
where we used the multiplication rule $\hat{r}_n\hat{t}_{\bf a}=\hat{t}_{R_n{\bf a}}\hat{r}_n$ of the space group generated by discrete $C_n$ rotations and translations of the lattice vectors, ${\bf e}_1$ and ${\bf e}_2$, and $R_n$ is a rotation operator by an angle $\frac{2\pi}{n}$ acting on a vector in the two-dimensional plane. 
Since the topological properties/charges of a disclination should not depend on what loops we choose, the topologically distinct disclinations for a fixed Frank angle lie in equivalence classes $[\bd{a}]$ of the holonomy group. 
The classification of disclination types then relies on identifying different conjugacy classes of the translation part ${\bf a}$, which is captured by the quotient 
$\hat{t}/[(R(\Omega)-1)\hat{t}]$ (this formula only applies for $\Omega\neq 0,2\pi$) \cite{benalcazar2014classification}. 
Note that this is a special case of the general statement that gauge fluxes are classified by conjugacy classes of the gauge group.

Let us construct these translation equivalence classes. For a given Frank angle, the number of conjugacy classes $[{\bf a}]^{(n)}$ of the translation piece is finite, and is in one-to-one correspondence with the different types of $C_n$ lattice rotation centers.  
For disclinations with Frank angle $\Omega=\pm\frac{\pi}{2}$, 
 $[{\bf a}]^{(4)}$ (we use the superscript $ ^{(n)}$ to denote the corresponding Frank angle $\frac{2\pi}{n}$) takes values in $\mathbb{Z}_2$ , i.e., $[{\bf a}]^{(4)}=0$ or $[{\bf a}]^{(4)}=1$, by which we represent the sum of the two components of $\mathbf{a}$ being even or odd respectively. 
In Fig.~\ref{fig:ConstructDis}(b,d), we show the examples of two different types of disclinations with Frank angle $\Omega=-\frac{\pi}{2}$ in a $C_4$ symmetric lattice. 
In Fig.~\ref{fig:ConstructDis}(b), choosing the closest loop to the disclination core (red loop with arrows), the holonomy is $\hat{t}_{-2\bd{e}_1}\hat{r}(-\frac{\pi}{2})$, leading to $[{\bf a}]^{(4)}=0$ while in Fig.~\ref{fig:ConstructDis}(d) the holonomy for the red loop with arrows is $\hat{t}_{-3\bd{e}_1}\hat{r}(-\frac{\pi}{2})$, leading to $[{\bf a}]^{(4)}=1$. 
Moving on, for a disclination of Frank angle $\Omega=\pi$, $[{\bf a}]^{(2)}$ takes values in $\mathbb{Z}_2\oplus\mathbb{Z}_2$ corresponding to the two components of the translation, $a_1, a_2$ being odd or even. When $[{\bf a}]^{(2)}=(0,0)$, the disclination core lies on a lattice site, when $[{\bf a}]^{(2)}=(0,1)$ or $[{\bf a}]^{(2)}=(1,0)$, the core is at the middle of the link between two lattice sites, and when $[{\bf a}]^{(2)}=(1,1)$, the core is located at the center of a plaquette. 
For a disclination of Frank angle $\Omega=\pm \frac{2\pi}{3}$, $[{\bf a}]^{(3)}$ is $\mathbb{Z}_3$ valued, corresponding to the modulo $3$ difference of two components of the translation holonomy, $(a_1-a_2) \mod 3$, being $0, 1$ or $-1$. When Frank angle $\Omega=\frac{2\pi}{6}$, the translation holonomy is zero, i.e., there is only one type of disclination. We label the only conjugacy class for this case as $[{\bf a}]^{(6)}=0$.
The configurations of all types of disclinations for each $C_n$ symmetry are shown in the Appendix \ref{sec: disclinations}. 

Since we will use it later, we also note that when a lattice has two disclinations $(\Omega_1,{\bf a}_1)$ and $(\Omega_2,{\bf a}_2)$, the total holonomy of a closed path enclosing both  disclination cores is $(\Omega_1+\Omega_2,{\bf a}_1+\hat{r}(\Omega_1){\bf a}_2)$\cite{PhysRevLett.111.047006,benalcazar2014classification}.

\section{Fractional disclination charge in TCIs with (generalized) Wannier representations}
\label{Sec:fractionaldisclinationcharge}
\subsection{Fractional disclination charge from Wannier orbital picture}
\label{Sec: fracdisWan}
\begin{figure}[!th]
\centering
\includegraphics[width=0.9\columnwidth]{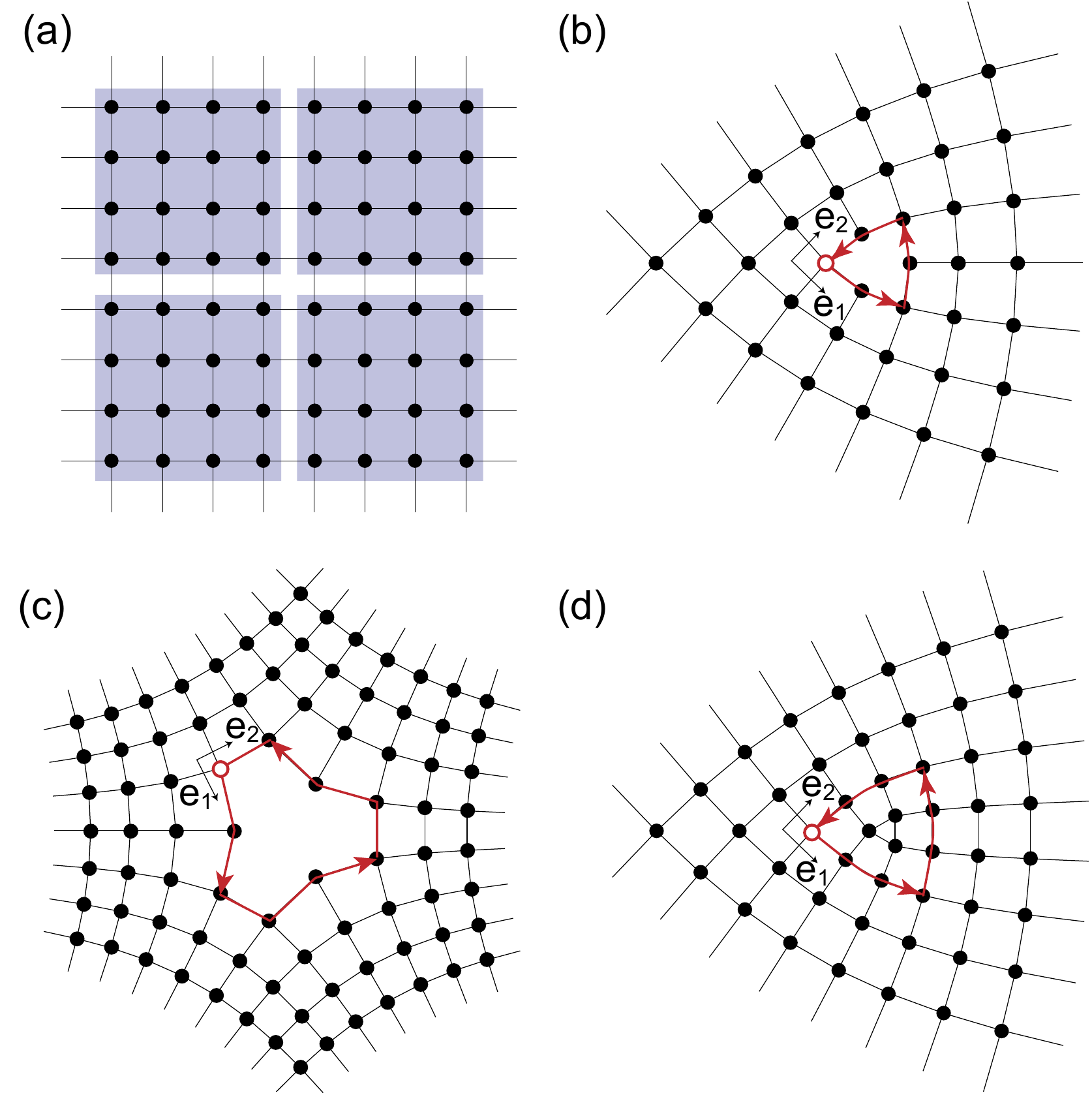}
\caption{(Color online) (a) A two-dimensional $C_4$ symmetric square lattice. Shadowed areas represent four sectors that are related by the $C_4$ rotation symmetry. (b)-(d) $C_4$ symmetric lattice with disclination of (b) $\Omega=-\frac{\pi}{2},[\bd{a}]^{(4)}=0$, (c) $\Omega=\pi,[\bd{a}]^{(2)}=(1,0)$ and (d) $\Omega=-\frac{\pi}{2},[\bd{a}]^{(4)}=1$. The red open circles indicate the starting points of the loop. See Appendix ~\ref{sec: disclinations} for plots of all types of disclinations for each $C_n$ symmetry.}
\label{fig:ConstructDis}
\end{figure}
\begin{figure}[t!]
\centering
\includegraphics[width=1\columnwidth]{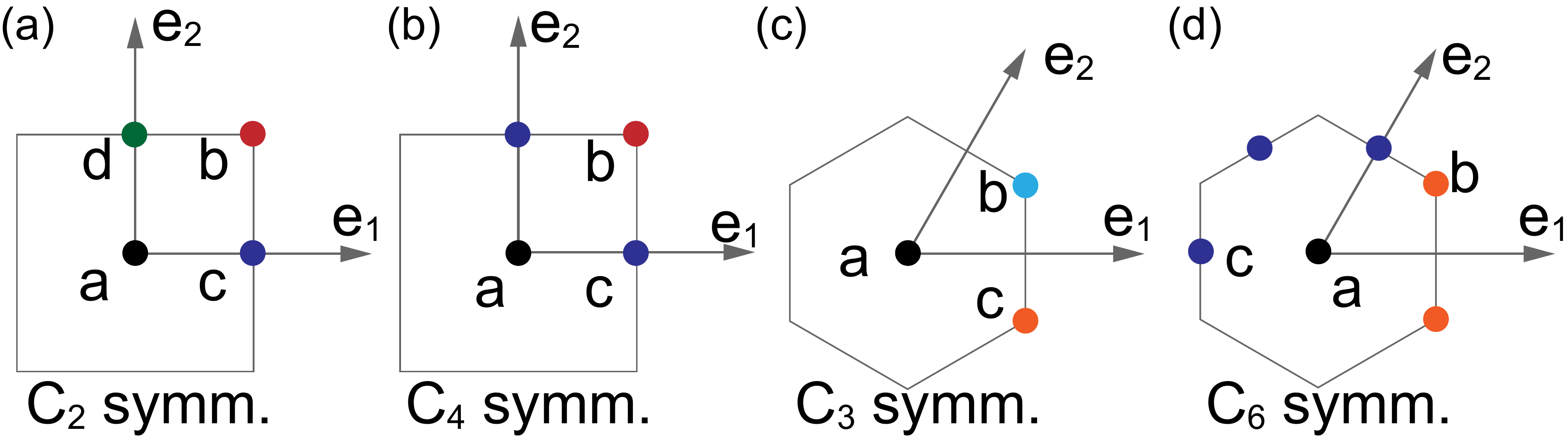}
\caption{(Color online) The unit cell, lattice vectors and maximal Wyckoff positions for (a) $C_2$, (b) $C_4$ symmetric TCIs with $\bd{a}_1=\hat{x},\bd{a}_2=\hat{y}$; (c) $C_3$ and (d) $C_6$ symmetric TCIs with $\bd{a}_1=\hat{x},\bd{a}_2=\frac{1}{2}\hat{x}+\frac{\sqrt{3}}{2}\hat{y}$.  In each unit cell, points of the same color belong to the same Wyckoff position and the number of points with same color indicates the multiplicity.}
\label{fig:WyckoffPos}
\end{figure}

If we can construct symmetric, spatially localized Wannier functions from the occupied Bloch states of a TCI, then we call the TCI an \emph{atomic insulator}. 
In real space, an atomic insulator can be viewed as a set of Wannier orbitals centered at high symmetry points, or symmetry-related points, called Wyckoff positions in each unit cell. 
In order to illustrate the Wyckoff positions in real space in more detail, let us denote $G$ as the entire (symmorphic) space symmetry group generated by $C_{n}$ rotation and lattice translations, and denote $G_{\alpha}\subset G$ as the site symmetry group of a Wyckoff position $\alpha$.
If we cannot find a finite group $H\subset G$ that satisfies $G_{\alpha}\subset H\subset G$, then we call the site symmetry group $G_{\alpha}$ a maximal subgroup of $G$ and 
Wyckoff position $\alpha$ a maximal Wyckoff position\cite{JenniferBuilding2018}.
In Fig.~\ref{fig:WyckoffPos}, we show the maximal Wyckoff positions (labeled by a letter) for each $C_n$ symmetry. 
In a $C_n$ symmetric TCI, a maximal Wyckoff position $\alpha$ is only invariant under the site symmetry group $C_{m_\alpha}, {m_\alpha}\leqslant n,$ and there are $\frac{n}{{m_\alpha}}$ points belonging to the same Wyckoff position per unit cell that are related by $C_n$ rotations. 
We denote $M_\alpha=\frac{n}{m_\alpha}$ as the multiplicity for the Wyckoff position $\alpha$. 
Minimal Wannier orbitals centered at a Wyckoff position $\alpha$ form irreducible representations (irreps) of the site symmetry group $C_{m_\alpha}$. We label such an irrep by its (total) angular momentum $l$, which is defined via the phase $e^{i\frac{2\pi l}{m_{\alpha}}}$ gained upon a $C_{m_\alpha}$ rotation, where $l=0,1\ldots m_{\alpha}-1$ for spinless case and $l=\frac{1}{2},\frac{3}{2}\ldots m_{\alpha}-\frac{1}{2}$ for spin-$\frac{1}{2}$ case. 
In order to have a $C_n$ symmetric configuration of Wannier orbitals in each unit cell, we need $M_\alpha$ Wannier orbitals with the same angular momentum $l$ located at each point belonging to the same Wyckoff position, say $\alpha$. 

One can induce a representation $\alpha_l$ of the space group generated by $C_{n}$ rotations and lattice translations from these $M_\alpha$ Wannier orbitals\cite{bradlyn2017topological,JenniferBuilding2018}.
Due to the lattice translation symmetry, it is convenient to understand the induced representation $\alpha_{l}$ in one unit cell in real space. Indeed, the induced representation $\alpha_{l}$ just represents an atomic insulator having Wannier orbitals with angular momentum $l$ located at Wyckoff position $\alpha$ in each unit cell throughout the entire lattice.
We denote this Wannier representation of an atomic insulator as $\sum_{\alpha,l}n_{\alpha}^{l}\alpha_{l}$,
where $n_\alpha^{l}\geqslant 0$ is the number of $\alpha_l$ representations, $l$ runs over all irreps of $C_{m_\alpha},$ and $\alpha$ runs over all Wyckoff positions. 

For example, in the $C_4$ symmetric case, $c_{0}$ means that there is one pair of Wannier orbitals with angular momentum $l=0$ located at the two equivalent Wyckoff positions labelled by $c$ (shown in Fig.~\ref{fig:WyckoffPos}(b)) in each unit cell, and $2c_{0}+a_{1}$ means that there are two pairs of Wannier orbitals with angular momentum $l=0$ located at two equivalent Wyckoff positions labelled by $c,$ plus a Wannier orbital with angular momentum $l=1$ located at the Wyckoff position labelled by $a$ in each unit cell.
In this sense, the set $\{n_{\alpha}^{l}\}$($\{n_{a}^{1}=1,n_{c}^{0}=2,\rm{others}=0\}$ for the above example) can be regarded as the configuration of Wannier orbitals in one unit cell, and it specifies a crystalline atomic insulator.  When the center of Wannier orbitals in one unit cell deviates from the center of the unit cell (position $a$ in Fig.~\ref{fig:WyckoffPos}),  the atomic insulator is said to be in an \emph{obstructed atomic limit}.

Since one Wannier orbital carries the charge of one electron, the spatial charge distribution in an atomic insulator is determined by the Wannier orbital configuration $\{n_{\alpha}^{l}\}$ as well.
We can generalize this idea to include fragile TCIs, in which an obstruction exists for constructing a set of symmetric Wannier functions (we will call this obstruction a ``Wannier obstruction" in the following for brevity)\cite{brouder2007exponential,soluyanov2011wannier}.
However, unlike a Chern insulator, which also has such a Wannier obstruction, when adding atomic insulator bands  to a fragile TCI, the Wannier obstruction can be resolved\cite{bradlyn2019disconnected,po2018fragile}, and the new combined TCI admits a Wannier orbital configuration. 
Consider the combination of the bands of a fragile TCI, and the bands of an atomic insulator $\{q_\alpha^l\},$ that admits a symmetric configuration of Wannier orbitals designated by $\{n_\alpha^l\}$. 
The charge distribution in this fragile TCI corresponds to a charge distribution where the Wannier orbital configuration $\{q_\alpha^l\}$ is taken away from the Wannier orbital configuration $\{n_\alpha^l\}$.
We can therefore define a generalized configuration of Wannier orbitals, $\{n^{l\prime}_{\alpha}=(n^l_\alpha-q^l_\alpha)\}$ with $n^{l\prime}_{\alpha}$ possibly being negative, to describe the charge distribution in this fragile TCI.

In order to derive the localized charges at disclinations in TCIs which admit a generalized Wannier orbital description, let us first establish the direct relation between the configuration of Wannier orbitals and the bulk polarization. We can illustrate this connection through the one-dimensional \emph{Su-Schrieffer-Heeger}(SSH) model\cite{SSHmodel}. Under inversion symmetry, the bulk polarization is quantized to be either $0$ or $\frac{e}{2}$ (we set the electronic charge to $e=1$ for simplicity from now on), corresponding to one Wannier orbital centered either at the center or at the boundary in each unit cell. Generalizing this correspondence to two dimensional lattices, the 2D coordinate inversion is equivalent to a $C_2$ rotation, and similar quantization conditions must hold for the components of the polarization vector. With $C_2$ symmetry we can consider the atomic insulator configurations. One Wannier orbital centered at Wyckoff position $c(d)$ in Fig.~\ref{fig:WyckoffPos} per unit cell corresponds to bulk polarization ${\bf P}=\frac{1}{2}{\bf e}_1$ (${\bf P}=\frac{1}{2}{\bf e}_2$), and one Wannier orbital located at Wyckoff position $b$ in Fig.~\ref{fig:WyckoffPos} per unit cell corresponds to bulk polarization ${\bf P}=\frac{1}{2}({\bf e}_1+{\bf e}_2)$. Other, non-maximal Wyckoff positions do not contribute to the polarization. Following this argument, we can repeat it for each $C_n$ symmetry and write down the bulk polarization in terms of the number of Wannier orbitals centered at each Wyckoff position:
\begin{align}
\label{eq:indexpolar}
{\bf P}^{(2)}&=\frac{n_b+n_c}{2}{\bf e}_1+\frac{n_b+n_d}{2}{\bf e}_2\ \rm{mod} \ 1,\nonumber\\
{\bf P}^{(4)}&=\frac{n_b+n_c}{2}({\bf e}_1+{\bf e}_2)\ \rm{mod} \ 1,\nonumber\\
{\bf P}^{(3)}&=\frac{(n_b-n_c)}{3}({\bf e}_1+{\bf e}_2)\ \rm{mod} \ 1,
\end{align}
where the superscript $(n)$ labels the $C_{n}$ symmetry and $n_{\alpha}=\sum_{l}n_{\alpha}^{l}$ for $\alpha=a,b,c,\cdots$. We note that $C_6$ symmetry forces the bulk polarization ${\bf P}^{(6)}$ to vanish, i.e., ${\bf P}^{(6)}=0$. This is because $C_{2}$ symmetry requires the polarization to be $0$ or $\frac{1}{2},$ and $C_{3}$ symmetry requires the polarization to be $0$, $\frac{1}{3}$ and $\frac{2}{3}$. Thus, in TCIs with $C_{6}$ symmetry which contains both $C_{2}$ and $C_{3}$ symmetry, the only possible polarization is $0$.

The work in Ref \onlinecite{benalcazar2019quantization} shows that insulators with trivial polarization may have quantized fractional charges that are exponentially localized at the corners of a finite two-dimensional lattice (or more generally in sectors subtending an angle of $2\pi/n$ for non-ideal geometries). Both the quanta and the localization of the fractional charge can be understood through the configuration of Wannier orbitals. We expect that the generalized Wannier orbital description captures the fractional charges at topological defects like disclinations in the same spirit. 
Provided that both the possible Wyckoff positions and disclination types are finite, we can do a case-by-case study to systematically investigate the charge fractionalization at each type of  disclination and for each $C_n$ symmetry. 

\begin{figure}[t!]
\centering
\includegraphics[width=1\columnwidth]{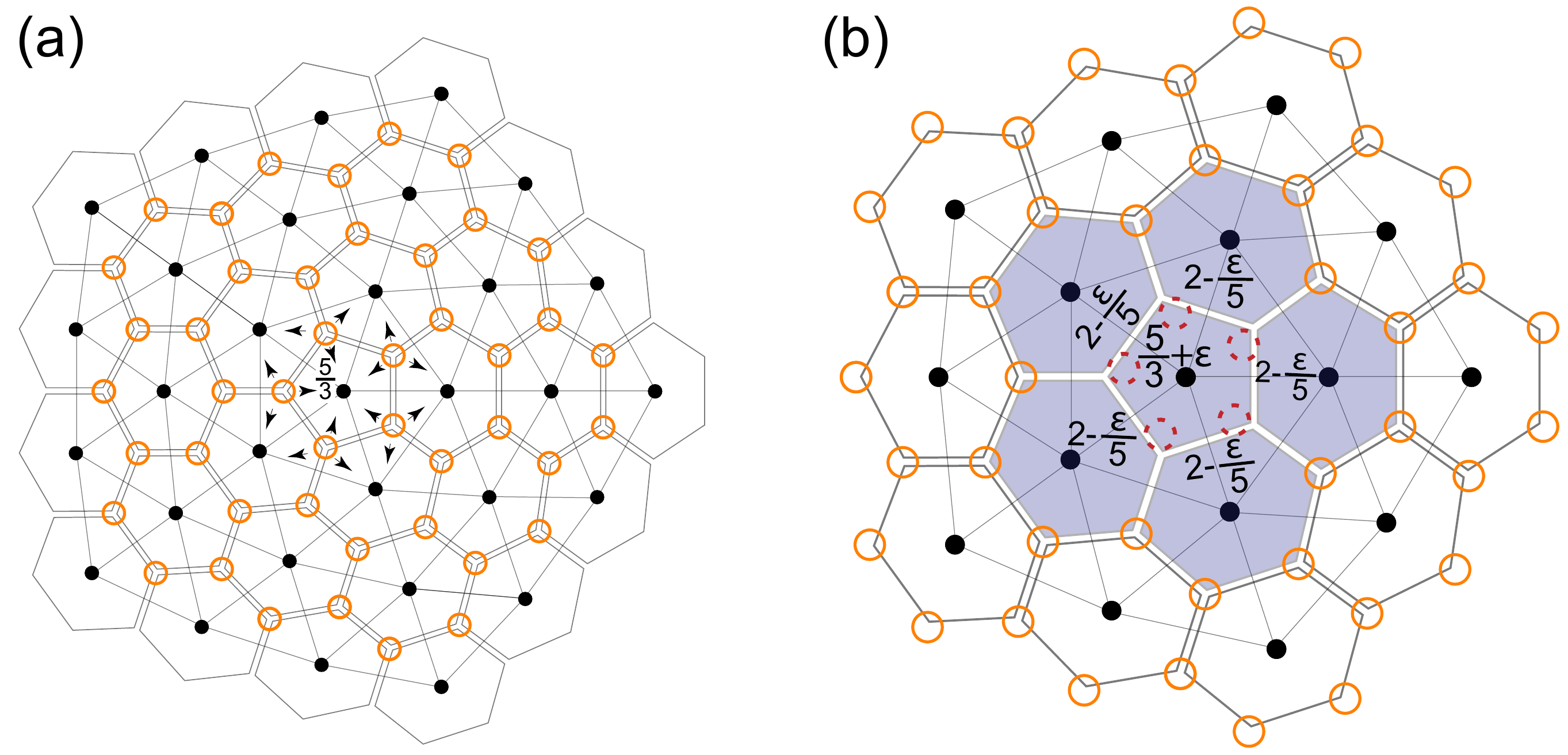}
\caption{
(Color online) 
The configuration of Wannier orbitals of a $C_6$ symmetric lattice with a disclination of $\Omega=-\frac{\pi}{3}$. Hexagons represent Wagner-Seitz primitive cells; black dots represent the triangular lattice sites; and solid lines between each black dot represent inter-cell hopping terms. Orange open circles represent Wannier orbitals located at Wyckoff position $b$ in Fig.~\ref{fig:WyckoffPos}(d). (a) The ideal case where all Wannier orbitals located at Wyckoff position $b$; (b) The general case where Wannier orbitals near the disclination core (dashed circles) shift slightly, and the charge distribution is distorted from the ideal case. However, the disclination still has a localized fractional charge. In order to clearly show the shifts of Wannier centers, we only show the nearest-neighboring and next nearest-neighboring unit cells of the disclination center.
}\label{fig:charge_c6}
\end{figure}

Let us take disclinations in $C_6$ symmetric lattices as an example to illustrate how we extract the fractional disclination charge from a configuration of Wannier orbitals. 
In order to be consistent with the definition of Wyckoff positions, we only remove or insert integer numbers of unit cells when constructing disclinations, $i.e.$, the lattice with disclinations preserves the same choice of unit cells as the perfect lattice.
In Fig.~\ref{fig:charge_c6}(a), we show a TCI with disclination in the zero-correlation length limit, where all Wannier orbitals are pinned at the Wyckoff position b. Each Wannier orbital contributes a charge of $\frac{1}{3}$ to its three nearest unit cells (as indicated by arrows in Fig.~\ref{fig:charge_c6}(a)). Then, we can clearly see that the unit cell at the core has fractional charge of $\frac{5}{3}$ and all the other unit cells have integer charge.  Therefore, a fractional disclination charge of $\frac{2}{3} \mod 1$ is localized at the disclination core.

In general, the disclination locally breaks the rotation and translation symmetry and the Wannier orbitals near the disclination deviate from the Wyckoff positions. We show an example in Fig.~\ref{fig:charge_c6}(b), where the Wannier orbitals closest to the core (dashed circles in Fig.~\ref{fig:charge_c6}(b)) shift slightly.
The shift of the Wannier orbitals causes changes in the charge distribution, and thus changes the charge in the unit cells at core and the $5$ unit cells closest to the core (shadowed area in Fig.~\ref{fig:charge_c6}(b)). The unit cell at the core now has fractional charge $\frac{5}{3}+\epsilon,$ and the 5 unit cells closest to the core now have fractional charge $2-\frac{\epsilon}{5}$. 
Other unit cells deep in the bulk and far away from the disclination center have integer charge. We can see that the fractional charge is still localized at the core and the deviation of Wannier orbitals only change its decay length.

In order to calculate the quantized fractional disclination charge, we need to enlarge the summation area -- we can sum up the charge over the unit cell at the core and the $5$ unit cells closest to the core.
By doing that, the change of charge in the unit cell at the core cancels the change of total charge over the closet $5$ unit cells around the core. The fractional disclination charge is still quantized to be $\frac{5}{3}$, which is the same as in the ideal case.
In general, for TCIs with larger correlation length, Wannier orbitals further away from the core may shift, which means the fractional disclination charge penetrates more into the bulk. However, we can always enlarge the summation area deeper into the bulk to calculate the quantized fractional charge.

It was previously shown that there are in-gap bound states appearing in lattices with disclinations \cite{PhysRevLett.111.047006,benalcazar2014classification,liu2018shift}. We emphasize that whether to fill these bound states or not does not change the fractional disclination charge, as long as all hopping terms are short-ranged. In the thermodynamic limit, edges are typically far away from the disclination core (further than the hopping range) and thus the in-gap states localized at the edges and the in-gap states localized at the disclination centers cannot be hybridized. Therefore, an in-gap state is either localized at the disclination core or localized at the edges. When filling an edge in-gap state, its contribution to the electronic charge at disclination center can be neglected. When filling an in-gap state localized at the disclination center, it only contributes an integer electronic charge. Thus, the fractional disclination charge will not change in both cases.

Following a similar counting method, we identify the fractional disclination charges for all possible configurations of Wannier orbitals at all types of disclinations. The results are summarized in Tables~\ref{tab:c2c4},\ref{tab:c3c6}.
\begin{table}[t]
\centering
\caption{Fractional charge at the core of different disclinations in $C_2$ and $C_4$ symmetric TCIs with zero Chern number.}
\label{tab:c2c4}
\renewcommand{\arraystretch}{1.5}
\begin{tabular}{p{0.08\columnwidth}|p{0.22\columnwidth}p{0.22\columnwidth}p{0.22\columnwidth}p{0.22\columnwidth}}
\hline\hline
\multirow{2}{*}{$C_2$} 
      & $\Omega=\pi$ & $\Omega=\pi$ & $\Omega=\pi$ 
      & $\Omega=\pi$ \\
      & $[{\bf a}]^{(2)}=(0,0)$ & $[{\bf a}]^{(2)}=(1,0)$ & $[{\bf a}]^{(2)}=(0,1)$ & $[{\bf a}]^{(2)}=(1,1)$\\
\hline
$Q_{dis}^{(2)}$ & $-\frac{n_{b}+n_{c}+n_{d}}{2}$ & $-\frac{n_{c}}{2}$ & $-\frac{n_{d}}{2}$ & $-\frac{n_{b}}{2}$\\
\hline\hline
\multirow{2}{*}{$C_4$} & $\Omega=\pm\frac{\pi}{2}$ & $\Omega=\pm\frac{\pi}{2}$    & $\Omega=\pi$     & $\Omega=\pi$\\                        & \multirow{2}{*}{$[{\bf a}]^{(4)}=0$} & \multirow{2}{*}{ $[{\bf a}]^{(4)}=1$} & 
$[{\bf a}]^{(2)}=(1,1)$  & $[{\bf a}]^{(2)}=(1,0)$\\
                       &                                      &                                       &
$[{\bf a}]^{(2)}=(0,0)$  & $[{\bf a}]^{(2)}=(0,1)$\\
\hline
$Q_{dis}^{(4)}$ & $\pm\frac{n_{b}+2n_{c}}{4}$ & $\mp \frac{n_{b}}{4}$ & $-\frac{n_{b}}{2}$ & $-\frac{n_{c}}{2}$\\ 
\hline\hline
\end{tabular}
\end{table}
\begin{table}[t]
\caption{Fractional charge at the core of different disclinations in $C_3$ and $C_6$ symmetric TCIs with zero Chern number. In $C_6$ symmetric TCIs, all types of disclinations for a given Frank angle have the same fractional charge.}
\label{tab:c3c6}
\renewcommand{\arraystretch}{1.5}
\begin{tabular}{p{0.1\columnwidth}|p{0.28\columnwidth}p{0.28\columnwidth}p{0.28\columnwidth}}
\hline\hline
\multirow{2}{*}{$C_3$} 
      & $\Omega=\pm\frac{2\pi}{3}$  & $\Omega=\pm\frac{2\pi}{3}$ & $\Omega=\pm\frac{2\pi}{3}$\\
      & $[{\bf a}]^{(3)}=\pm 1$ & $[{\bf a}]^{(3)}=\mp1$ & $[{\bf a}]^{(3)}=0$\\ 
\hline
$Q_{dis}^{(3)}$ & $\mp \frac{n_{b}}{3}$ &  $\mp \frac{n_{c}}{3}$ & $\pm\frac{n_{b}+n_{c}}{3}$\\ 
\hline\hline
$C_6$
      & $\Omega=\pm\frac{\pi}{3}$ & $\Omega=\pm\frac{\pi}{3}$
      & $\Omega=\pm\frac{2\pi}{3}$ \\
    %   & ${\bf a}=(0,0)$ & ${\bf a}=(0,0)$ & ${\bf a}=(0,0)$\\
\hline
$Q_{dis}^{(6)}$ & $\pm\frac{2n_{b}+3n_{c}}{6}$ & $\pm \frac{2n_{b}+3n_{c}}{3}$ & $\pm \frac{2n_{b}+3n_{c}}{2}$\\ 
\hline\hline
\end{tabular}
\end{table}
As mentioned above, based on this data we can connect the fractional disclination charges to the generalized Wannier configuration. The disclination charge can generically be divided into two parts. One part is associated with the rotation part of the holonomy, corresponding to how many $\frac{2\pi}{n}$ sectors are taken off or inserted. The other part is associated with the translation piece of the holonomy, and corresponds to the fractional charge at dislocations generated by nonzero bulk polarization\cite{CohPolarChern,Randislocmode,DislocationPiFlux,Bulk-defect,DislocationCharges}. 
In order to simplify the notation, we denote ${\bf T}^{(n)}=a_1{\bf d}_2-a_2{\bf d}_1$ as the vector perpendicular to the translation part of the holonomy, where ${\bf d}_i\cdot {\bf e}_j=\delta_{ij}$ is the basis of the dual space.
According to the addition rule of disclination holonomies $(\Omega_1,{\bf a}_1)+(\Omega_2,{\bf a}_2)=(\Omega_1+\Omega_2,{\bf a}_1+\hat{r}(\Omega_1){\bf a}_2)$, the results in Table ~\ref{tab:c2c4},\ref{tab:c3c6} can be summarized in the following index equations:
\begin{align}
\label{eq:discharge}
Q_{dis}^{(2)}&=\frac{\Omega}{2\pi}(n_b+n_c+n_d)+{\mathbf {T}^{(2)}\cdot \mathbf{P}^{(2)}}\mod 1,\nonumber\\
Q_{dis}^{(4)}&=\frac{\Omega}{2\pi}(n_b+2n_c)+{\mathbf{T}^{(4)}\cdot \mathbf{P}^{(4)}}\mod 1,\nonumber\\
Q_{dis}^{(3)}&=\frac{\Omega}{2\pi}(n_b+n_c)+{\mathbf{T}^{(3)}\cdot \mathbf{P}^{(3)}} \mod 1,\nonumber\\
Q_{dis}^{(6)}&=\frac{\Omega}{2\pi}(2n_b+3n_c) \mod 1.
\end{align}
Although for a fixed $\Omega$ the vector $\bd{T}^{(n)}$ may change for different choices of loops, the term $\bd{T}^{(n)}\cdot \bd{P}^{(n)}$ gives the same fractional values for equivalent disclination types.
Since all $n_\alpha$'s in Eq.~\eqref{eq:discharge} are integer, we find, in general, the disclination charge is fractionalized in units of $\frac{1}{n}$ for $C_n$ symmetric TCIs. 
Additionally, for spin-$\frac{1}{2}$ TCIs, at each Wyckoff position, the Wannier orbitals form a set of Kramers' pairs in the presence of TRS. In that case, the number of orbitals at each point of a Wyckoff position $\alpha$ must be even, leading to  fractional disclination charge quantized in units of $\frac{2}{n}$.

\subsection{Topological indices for fractional disclination charge in terms of rotation invariants}
\label{sec:fracdisrotidx}
While it is straightforward to extract the fractional disclination charge pictorially from the configuration of Wannier orbitals $\{n_{\alpha}^{l}\}$, it is not straightforward to obtain the configuration of Wannier orbitals for a given Bloch Hamiltonian $h(\bd{k})$. 
In this section, we utilize band representation theory \cite{bradlyn2017topological,po2017symmetry,JenniferBuilding2018,bradlyn2019disconnected,kruthoff2017topological} to link the configuration of Wannier orbitals $\{n_{\alpha}^{l}\}$ in real space to the rotation invariants in $\bd{k}$ space. A benefit of rotation invariants is  that they can be readily extracted from first principle calculations for real materials, or tight binding models. Then we derive the indices for fractional disclination charge in terms of the rotation invariants for both spin-$\frac{1}{2}$ and spinless atomic insulators and fragile TCIs. 

In the presence of the $C_n$ rotation symmetry, the Bloch Hamiltonian $h(\mathbf{k})$ satisfies $\hat{r}_{n}h(\mathbf{k})\hat{r}_{n}^{\dag}=h(R_{n}\mathbf{k})$, where $\hat{r}_{n}$ is the n-fold rotation operator and $R_{n}$ is a n-fold rotation acting on crystal momentum $\bd{k}$. 
At the HSP $\bd{\Pi}^{(n)}$ that satisfies $\bd{R}_n\bd{\Pi}^{(n)}=\bd{\Pi}^{(n)}$ modulo a reciprocal lattice vector, the Hamiltonian commutes with the rotation operator, $[\hat{r}_n,h(\bd{\Pi}^{(n)})]=0$. 
Therefore, given the eigenstates of the Hamiltonian at $\bd{\Pi}^{(n)}$, $\ket{u^{m}(\bd{\Pi}^{(n)})}$, we can calculate the eigenvalues of rotation operator $\hat{r}_n$ at $\bd{\Pi}^{(n)}$ by diagonalizing the matrix $\bra{u^{l}(\bd{\Pi}^{(n)})} \hat{r}_n\ket{u^{m}(\bd{\Pi}^{(n)})}$, where $l$ and $m$ run over the occupied bands. 
For spinless TCIs, $\hat{r}_n^n=1$ and we denote $\Pi_p^{(n)}=e^{i\frac{2\pi i(p-1)}{n}}, p=1,2,\ldots,n$ as the eigenvalues of $\hat{r}_n$ at $\bd{\Pi}^{(n)}$. 
For spin-$\frac{1}{2}$ TCIs, $\hat{r}_n^n=-1$ and $\hat{r}_{n}=\hat{r}_{n}^{L}\otimes e^{-i\frac{\pi}{n}\sigma_{z}}$, where $\hat{r}^L_n$ is a $n-$fold rotation on  the orbital degrees of freedom, and $e^{-i\frac{\pi}{n}\sigma_{z}}$ rotates the spin by $\frac{2\pi}{n}$. 
We denote the eigenvalues of $\hat{r}_n$ at $\bd{\Pi}^{(n)}$,  in this case, as $\Pi^{(n)}_{p}=e^{\frac{2\pi i(p-\frac{1}{2})}{n}}, p=1,2,\ldots,n$. We define the rotation invariants as
\begin{align}
\label{eq:rotind}
[\Pi^{(n)}_p]=\#\Pi^{(n)}_p-\#\Gamma^{(n)}_p,
\end{align}
where $\#\Pi^{(n)}_p$ is the number of the occupied bands with the eigenvalue $\Pi^{(n)}_p$ for both spinless TCIs and spin-$\frac{1}{2}$ TCIs.

Following the method in Refs \onlinecite{bradlyn2017topological,po2017symmetry,JenniferBuilding2018,bradlyn2019disconnected,kruthoff2017topological}, one can induce a band representation of the space group, which is generated by $C_n$ rotation and lattice translation, from a configuration of Wannier orbitals at Wyckoff positions $\{n_{\alpha}^{l}\}$, and then read off the eigenvalues of the rotation operators at HSPs in BZ. This method allows us to connect a configuration of Wannier orbitals in real space to the symmetry representations at HSPs in ${\bf k-}$space. We defer the detailed derivation of such a process for all configurations of Wannier orbitals in $C_2, C_3,C_4$
 and $C_6$ symmetric TCIs to Appendix.~\ref{sec: BandRep}.
Let us denote $l_{p\alpha}^{\Pi^{(n)}}$ as the number of the $p-$th eigenvalue of $\hat{r}_n$ at the HSPs $\Pi^{(n)}$ for the band representation $\alpha_l$ induced from the Wannier representation of Wannier orbitals with angular momentum $l$ at the Wyckoff position $\alpha$.
Consider a TCI with the Wannier orbital configuration $\{n_\alpha^{l}\}$ ($n_\alpha^{l}$ can be negative for fragile TCIs), the $\#\Pi^{(n)}_{p}$ can be calculated by summing over the contributions from band representations induced from all $\alpha_l$'s. 
Therefore, for given Bloch bands with certain $\#\Pi_p^{(n)}$ at each HSP, one can get a series of equations,
\begin{align}
\sum_{\alpha,l} n_\alpha^{l} l_{p\alpha}^{\Pi^{(n)}}=\#\Pi_p^{(n)},
\label{eq:linearrot}
\end{align}
involving the configuration of Wannier orbitals $\{n_\alpha^{l}\}$.
Additionally, for an insulator with a gapped energy spectrum, the number of occupied bands $\nu$ is constant across the BZ,
which imposes a constraint on the total number of Wannier orbitals,
\begin{align} 
\sum_{\alpha,l}M_{\alpha} n_\alpha^{l}=\nu=\sum_p\#\Pi_p^{(n)}.
\label{eq:filling}
\end{align} 
Recall that $M_{\alpha}$ is the multiplicity of a given Wyckoff position $\alpha$.
Combining Eq.~\eqref{eq:linearrot} and Eq.~\eqref{eq:filling}, one can get a set of linearly independent equations, from which we find the integers, $\{n_{\alpha}=\sum_{l}n_{\alpha}^{l}\}$ in terms of the rotation invariants.

Let us take $C_{6}$ symmetric spinless TCIs as an example. 
Given the number of occupied bands $\nu$, at the HSP $\bd{\Pi}^{(n)}$, we only have $n-1$ independent $\#\Pi^{(n)}_{p}$s due to Eq.~\eqref{eq:filling}. We choose the set of independent numbers for each HSP in the BZ as $\left(\#\Gamma^{(6)}_{1,2,3,4,5}, \#K^{(3)}_{1,2}, \#M^{(2)}_{1}\right)$. 
Meanwhile, Eq.\eqref{eq:filling} also leads to constraints on the rotation invariants defined in Eq.\eqref{eq:rotind}:
\begin{equation}
\label{eq: con_rotinv}
\sum_{p}[\Pi^{(n)}_p]=0.
\end{equation}
For the $C_{6}$ symmetric case, there are 5 nonzero rotation invariants $\bigg(\left[K^{(3)}_{1}\right],\left[K^{(3)}_{2}\right],\left[K^{(3)}_{3}\right],\left[M^{(2)}_{1}\right],\left[M^{(2)}_{2}\right]\bigg),$ and only three of them are independent due to Eq.\eqref{eq: con_rotinv}.
We choose to use $\bigg(\left[K^{(3)}_{1}\right],\left[K^{(3)}_{2}\right],\left[M^{(2)}_{1}\right]\bigg)$. 
With 8 independent $\#\Pi^{(n)}_{p}$'s, and the number of occupied bands $\nu$, we can get 8 equations from Eq.~\eqref{eq:linearrot}, and one equation from Eq.~\eqref{eq:filling} for $\{n_\alpha^{l}\}$ , all of which are linearly independent.
With $n_{\alpha}=\sum_{l}n_\alpha^{l}$, we can get 9 linearly independent equations for 11 unknowns $(n_{a},n_{b},n_{c},n_{a}^{0},n_{a}^{1},n_{a}^{2},n_{a}^{3},n_{a}^{4},n_{b}^{0},n_{b}^{1},n_{c}^{0})$
Since we have more unknowns than equations, we cannot get unique solutions. However, according to Eq.~\eqref{eq:discharge}, the fractional charge in a disclination only depends on $n_{b}$ and $n_{c}$. Thus, we can  write down $n_{b}$ and $n_{c}$ in terms of the 3 independent rotation invariants and 2 unknowns (we choose $n_{c}^{0}$ and $n_{b}^{1}$ here),
\begin{equation}
\label{eq: c6numer}
\begin{array}{l}
n_{b}=\left[K_{2}^{(3)}\right]+3n_{b}^{1},
\\
\\
n_{c}=\frac{1}{2}\left[M^{(2)}_{1}\right]+2n_{c}^{0}.
\end{array}
\end{equation}
This is a crucial result. Note that for obstructed atomic insulators or fragile TCIs, $n_\alpha^l$s must be integers, either positive or negative. Thus, the undetermined part of $n_b$ and $n_c$ are a multiple of 3 and a multiple of 2 respectively, which will \emph{not} affect the fractional charge according to Eq.~\eqref{eq:discharge}. Substituting Eq.~\eqref{eq: c6numer} into Eq.~\eqref{eq:discharge}, we can get an index for the fractional disclination charge \emph{solely} in terms of rotation invariants
\begin{equation}
\label{eq:c6charge}
Q_{dis}^{(6)}=\frac{\Omega}{2\pi}\left(2\left[K_{2}^{(3)}\right]+\frac{3}{2}\left[M^{(2)}_{1}\right]\right)\mod 1.
\end{equation}
For the spin-$\frac{1}{2}$ case, we only need to replace the integer superscript $l=0,1,\cdots, m_{\alpha}-1$ in $n_{\alpha}^{l}$ by half integer $l=\frac{1}{2},\frac{3}{2},\cdots,m_{\alpha}-\frac{1}{2}$ and then we recover exactly the same equation set, and thus the same index, i.e., Eq.~\eqref{eq:c6charge} is applicable for both the spinless case and spin-$\frac{1}{2}$ case.

For $C_{2}$, $C_{3},$ and $C_{4}$ symmetric insulators, we can do the same thing and solve for all $n_{\alpha}$'s needed to calculate the fractional charge in Eq.~\eqref{eq:discharge}. We find that, for each $C_n$ symmetry, the unknown parts in the solutions of $n_{\alpha}$'s \emph{never} contribute to the fractional charges. Finally, we can write down the fractional disclination charge indices in terms of rotation invariants as
\begin{equation}
\label{eq:dischargek}
\begin{array}{l}
\begin{aligned}
Q_{dis}^{(2)}=&\frac{\Omega}{2\pi}\left(\frac{1}{2}\left[X^{(2)}_{1}\right]+\frac{1}{2}\left[Y^{(2)}_{1}\right]+\frac{1}{2}\left[M^{(2)}_{1}\right]\right)
\\
&+\mathbf{T}^{(2)}\cdot\mathbf{P}^{(2)} \ \rm{mod} \ 1,
\end{aligned}
\\
\\
\begin{aligned}
Q_{dis}^{(3)}=\frac{\Omega}{2\pi}\left(\left[K^{(3)}_{2}\right]+\left[K^{\prime(3)}_{2}\right]\right)+\mathbf{T}^{(3)}\cdot\mathbf{P}^{(3)} \ \rm{mod} \ 1,
\end{aligned}
\\
\\
\begin{aligned}
Q_{dis}^{(4)}=&\frac{\Omega}{2\pi}\left(\left[X^{(2)}_{1}\right]+\frac{3}{2}\left[M_{3}^{(4)}\right]-\frac{1}{2}\left[M_{1}^{(4)}\right]\right)
\\
&+\mathbf{T}^{(4)}\cdot\mathbf{P}^{(4)}\ \rm{mod} \ 1.
\end{aligned}
\end{array}
\end{equation}
Following Eq.~\eqref{eq:indexpolar}, the bulk polarization can be written down as
\begin{equation}
\begin{array}{l}
 \mathbf{P}^{(2)}=\frac{1}{2}\left(\left[X^{(2)}_{1}\right]\mathbf{e}_{1}+\left[Y^{(2)}_{1}\right]\mathbf{e}_{2}\right),
 \\
 \\
 \begin{aligned}
 \mathbf{P}^{(3)}=\frac{1}{3}\bigg(&\left[K^{(3)}_{2}\right]-\left[K^{\prime(3)}_{2}\right]
 \\&+2\left[K^{(3)}_{1}\right]-2\left[K^{\prime(3)}_{1}\right]\bigg)\left(\mathbf{e}_{1}+\mathbf{e}_{2}\right),
 \end{aligned}
 \\
\\
 \mathbf{P}^{(4)}=\frac{1}{2}\left(\left[X^{(2)}_{1}\right]+\left[M_{3}^{(4)}\right]-\left[M_{1}^{(4)}\right]\right)\left(\mathbf{e}_{1}+\mathbf{e}_{2}\right).
 \label{eq:kpolar}
 \end{array}
 \end{equation}
Detailed derivations of Eq.~\eqref{eq:dischargek} and Eq.~\eqref{eq:kpolar} are shown in Appendix~\ref{sec: BandRep}. These indices capture the fractional portions of charges that are exponentially localized at the disclination. They are robust against perturbations that preserve the $C_n$ symmetry and bulk energy gap.

As discussed above, the indices in Eq.\eqref{eq:c6charge} and Eq. \eqref{eq:dischargek} apply for both the spinless and spin-$\frac{1}{2}$ $C_n$ symmetric TCIs, as long as the TCIs admit a (generalized) Wannier description. 
Therefore, the invariants in Eq.~\eqref{eq:dischargek} are subject to the constraints imposed by a vanishing Chern number, since TCIs with non-zero Chern number do not admit a (generalized) Wannier representation\cite{Thonhauser2006WannierChern}.
For the disclinations with the minimal Frank angle $\Omega=\frac{2\pi}{n}$, the coefficient in front of the parenthesis in Eq.~\eqref{eq:dischargek} is $\frac{\Omega}{2\pi}=\frac{1}{n}$. 
Since the term coupled to $\frac{\Omega}{2\pi}$ is an integer in Eq.~\eqref{eq:dischargek}, the minimal fractional disclination charge in $C_3$ symmetric TCIs is $\frac{1}{3}$.
However, it is not obvious from the indices themselves whether this is the same for $C_2, C_4,$ and $C_6$ symmetric TCIs.
Using the results in Ref. \onlinecite{fang2012bulk},
 we find that the Chern numbers in $C_2,C_4,$ and $C_6$ symmetric TCIs satisfy, 
\begin{align}
\label{eq: Chern_rot}
ch^{(2)}&=[X_1^{(2)}]+[Y_1^{(2)}]+[M_1^{(2)}] \mod 2,\nonumber\\
ch^{(4)}&=[M_1^{(4)}]+2[M_2^{(4)}]+3[M_3^{(4)}]+2[X_1^{(2)}]\mod 4,
\nonumber\\
ch^{(6)}&=3[M_2^{(2)}]+4[K_2^{(3)}]+2[K_3^{(3)}] \mod 6.
\end{align}
Imposing the vanishing Chern number constraints on the rotation invariants, we find the term coupled to $\frac{\Omega}{2\pi}$ must also be an integer for $C_2,C_4,$ and $C_6$ TCIs.
Therefore from the indices, we find that the disclination charge is fractionalized in units of $\frac{1}{n}$ a $C_n$ symmetric TCIs. 

Now let us discuss the influence of TRS on this quantization.
TRS requires that the eigenvalues of rotation operators at time reversal invariant HSPs must come as complex conjugate pairs if they are complex. 
For spinless TCIs with $\hat{T}^2=1$ ($\hat{T}$ is the time reversal operator), TRS leads to $[M_2^{(4)}]=[M_4^{(4)}]$ for $C_4$ symmetry, $[K^{\prime(3)}_2]=[K^{\prime(3)}_3],[K^{\prime(3)}_3]=[K^{\prime(3)}_2]$ for $C_3$ symmetry, and $[K^{(3)}_2]=[K^{\prime(3)}_3]$ for $C_6$ symmetry.
These constraints do not further influence the fractionalization of disclination charge from the case without TRS. 

For the spin-$\frac{1}{2}$ case TCIs have $\hat{T}^2=-1$, and TRS imposes more constraints on rotation invariants.
First, for rotation invariants at all two-fold HSPs ${\bf \Pi}^{(n)}$, $[\Pi_1^{(2)}]=[\Pi_2^{(2)}]$, since the conservation of the number of occupied bands over each HSP requires that $[\Pi_1^{(2)}]+[\Pi_2^{(2)}]=0$; hence all two-fold rotation invariants are trivial $[\Pi^{(2)}_i]=0,i=1,2$. 
This directly leads to integer disclination charge in $C_2$ symmetric spin-$\frac{1}{2}$ TCIs with TRS.
Similarly, for $Q_{dis}^{(6)}$, the term proportional to $\frac{\Omega}{2\pi}$ is $2[K_2^{(3)}]$, meaning that the minimal fractional disclination charge for $C_6$ symmetric spin-$\frac{1}{2}$ TCIs with $T^2=-1$ is $\frac{2}{6}=\frac{1}{3}$. 
Second, for $C_4$ symmetric spin-$\frac{1}{2}$ TCIs, TRS requires that $[M_1^{(4)}]=[M_4^{(4)}], [M_2^{(4)}]=[M_3^{(4)}]$.
Combining the constraints from conservation of the number of occupied bands $[M_1^{(4)}]+[M_2^{(4)}]+[M_3^{(4)}]+[M_4^{(4)}]=0$, we find $[M_1^{(4)}]=-[M_3^{(4)}]$. So the term proportional to $\frac{\Omega^{(4)}}{2\pi}$ in Eq.~\eqref{eq:dischargek} must be an even number. This means that the minimal fractional disclination charge in $C_4$ spin-$\frac{1}{2}$ TCIs with addtional TRS is $\frac{1}{2}$. 
In summary, with additional TRS, the disclination charge is fractionalized in units of $\frac{2}{n}$ in $C_n$ symmetric spin-$\frac{1}{2}$ TCIs 
\bibnote{Note that in $C_3$ symmetric TCIs, for the fractionalization unit, $\frac{1}{3}$ is equivalent to $\frac{2}{3}$, since $\frac{2}{3}+\frac{2}{3}=\frac{1}{3}\mod 1$.}.

Apart from the doubling of the fractional quantization of disclination charge, TRS in spin-$\frac{1}{2}$ TCIs implies that \emph{integer} disclination charge can still be a robust, symmetry-protected topological feature. 
This is because any changes due to time-reversal symmetric perturbations come as Kramers' pairs, which carry charge $2$.
Therefore, one can only remove or insert an even number of electrons to the disclination while preserving TRS.
In order to change the charge from even to odd, either a topological phase transition must happen or TRS must be broken.
Thus, we find that the amount of topologically robust charge can be well-defined modulo $2$ in this case instead of modulo $1$. 
When the disclination carries an odd-integer charge, TRS implies a form of 2D spin-charge separation\cite{ran2008,qi2008b} that is a generalization of the spin-charge separation at the domain wall of the spinful SSH model\cite{SSHmodel}.

\begin{figure}[t]
\centering
\includegraphics[width=\columnwidth]{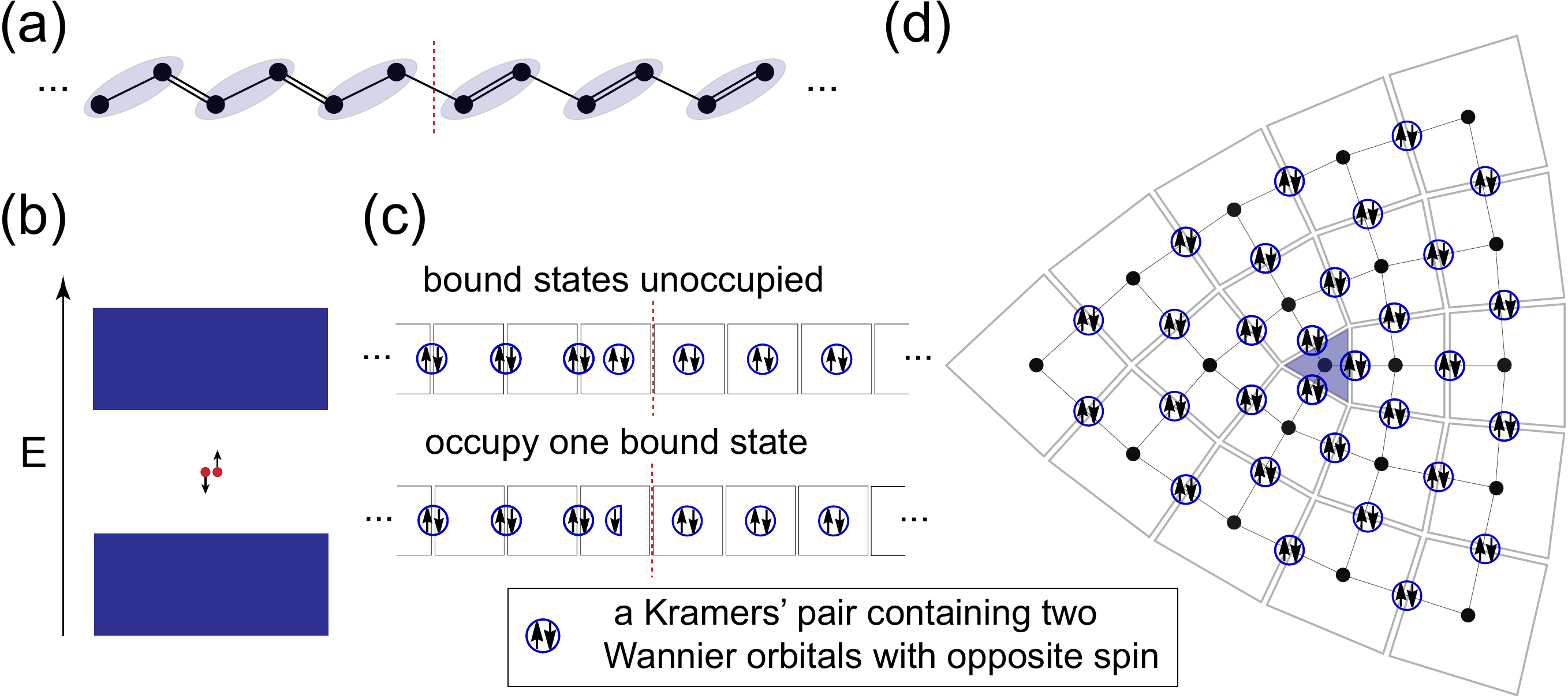}
\caption{(Color online)
(a)Lattice configuration of an infinite SSH chain with a domain wall in the middle. The red dashed line indicates the domain wall (b) The schematic energy spectrum of the SSH model with the domain wall. (c) Wannier orbital configurations of two different fillings, both mid-gap modes empty (top), or only one filled (bottom). (d) A time reversal invariant Wannier configuration in a $C_4$ symmetric lattice with a disclination of $\Omega=-\frac{\pi}{4}$, $[{\bf a}]^{(4)}=0$. The shaded unit cell indicates the disclination core. There is an odd charge but zero spin at the disclination core.
}
\label{fig:SpinChargeSep}
\end{figure}

Let us first illustrate how to use a Wannier orbital picture to understand the spin-charge separation at the domain wall of a spinfull SSH model. 
Fig.~\ref{fig:SpinChargeSep}(a) shows the lattice configuration of an infinite 1D  SSH chain.
There are two sites in one unit cell and each of them can contain two electrons, one with spin up and one with spin down.
The single (double) bounds represent hoppings with weaker (stronger) amplitudes. The left part of the lattice is in the topological phase and the right part of the lattice is in the trivial phase, creating a domain wall (dashed line) in the middle. 
When imposing TRS with $T^{2}=-1,$ which acts locally in real space, two Wannier orbitals are grouped together to form a Kramers' pair, which carries $2$ electric charges and zero spin. Without loss of generality, we assume one Wannier orbital has $S_{z}=+\frac{1}{2},$ and its Kramers' partner has $S_{z}=-\frac{1}{2}$ (represented by arrows pointing up and down). 

In the left part (topological phase), the Kramers' pairs are localized in between unit cells. For each Kramers' pair, the Wannier orbital with $S_{z}=+\frac{1}{2}$ contributes $\frac{1}{2}$ charge to each of the two adjacent unit cells, as does the Wannier orbital with $S_{z}=-\frac{1}{2}$. As such, they contribute a charge of $1,$ and zero spin to each of the two adjacent unit cells.
In the right part (trivial phase), the Kramers' pairs are localized at the center of the unit cells, and they contribute a charge of $2$ and zero spin to the unit cells in which they reside.

Figure~\ref{fig:SpinChargeSep}(b) shows a schematic energy spectrum of such a configuration, where two mid-gap states (red dots) are induced at the domain wall. 
These two mid-gap states are degenerate, and the degeneracy is protected by TRS.
If both midgap states are filled then the corresponding Wannier orbital configuration is shown in Fig~\ref{fig:SpinChargeSep}(c). Following the counting method discussed above, each unit cell away from the domain wall has a charge of $2$ and zero spin.
However, a unit cell at  the domain wall (to the left to be precise) carries a charge of $3$ and zero spin.
Note that if the two midgap states are empty, the charge at the domain wall changes by $2$, i.e., it still remains odd, and the spin remains zero, i.e., it remains time-reversal invariant.
However, if we slightly break TRS, the two mid-gap states are split and become spin polarized in some direction (we will call it the $z$-direction without loss of generality). Suppose we fill the low-energy state with $S_z=-\frac{1}{2}$, then the domain wall carries a charge of $2$ and a net spin $S_z=-\frac{1}{2}$. 
Thus, in the SSH model, we find the spin and charge degrees of freedom can be separated. From the Wannier orbital picture, it is clear to see that although the domain wall has odd charge, all the Wannier orbitals can be grouped in Kramers' pairs and thus are invariant under time reversal. On the other hand, when the domain wall has zero net charge, an unpaired spin appears. Time reversal flips the spin, and thus the Wannier configuration breaks TRS.

By analogy we show an example of an ideal, time reversal invariant configuration of Wannier orbitals that carries odd-integer charge, but no spin at a disclination.
As shown in Fig.~\ref{fig:SpinChargeSep}(d), the Wannier orbitals with $S_{z}=+\frac{1}{2}$ in the three Kramers' pairs around the disclination core contribute a total charge of $3/2,$ and the three Wannier orbitals with $S_{z}=-\frac{1}{2}$ contribute a total charge of $3/2$ to the disclination core. In total, we have a charge of $3$ and zero spin at the core, which is a manifestation of spin-charge separation.
The odd integer disclination charge is robust upon the addition of Wannier orbitals of Kramers' pairs to the disclination core since such addition only changes the disclination charge by $2$. If we want to change the odd charge to even charge, we can add a single electron to the core. However, a single electron introduces an un-paired spin to the disclination core, and indicates broken TRS.

Despite the clear distinction between even and odd integer charges,  the disclination indices, written \emph{solely} in terms of rotation invariants \emph{cannot} distinguish the evenness or oddness of the disclination charge. 
The reason is that when solving for $n_\alpha$ from rotation eigenvalues at HSPs with the constraints of TRS, the undetermined part is not generally an \emph{even} number.  
Let us elaborate this in $C_{6}$ symmetric case: the undetermined part $n^{\frac{1}{2}}_{c}$ of $n_{c}$ in Eq.~\eqref{eq: c6numer} can be an odd integer
\bibnote{For spin-$\frac{1}{2}$ TCIs, the sets of equations between $n_\alpha^l$ and $\Pi_i^{(n)}$ remain the same only with the replacement of $l=0,1\ldots, m_\alpha-1 $ by $l=\frac{1}{2},\frac{3}{2},\ldots m_\alpha -\frac{1}{2}$.},
resulting in the contribution of an odd disclination charge according to Eq.~\eqref{eq:discharge}. Thus the undetermined parts $n^{\frac{1}{2}}_c$, cannot be dropped when substituting Eq.~\eqref{eq: c6numer} into Eq.~\eqref{eq:discharge} if the disclination charge is defined modulo 2.
This implies that the rotation invariants at HSPs alone \emph{cannot} distinguish the TCIs with topologically non-trivial odd disclination charges from the TCIs with trivial even disclination charges. Therefore the symmetry representation at HSPs can not fully classify the spin$-\frac{1}{2}$ TCIs with TRS, which is consistent with a recent work of Ref. \onlinecite{bradlyn2019disconnected}.

It is clear from Eq. \eqref{eq:c6charge} and Eq. \eqref{eq:dischargek} that the fractional charges in all types of disclinations for two-dimensional $C_n$ symmetric TCIs that are either obstructed atomic insulators or fragile topological insulators are completely determined by the rotation invariants. Let us discuss an important consequence of this result.
Although the rotation invariants are sufficient to determine the charge response at the disclinations, given all rotation invariants, it is not enough information to determine whether the TCI is a fragile TCI or an obstructed atomic insulator. 
For example, when there is $C_{2}$ symmetry, a fragile TCI represented as 
$b_{0}+c_{0}+d_{0}+a_{1}-a_{0},$ and an obstructed atomic insulator represented as $b_{0}+c_{0}+d_{0}$, both of which have 3 occupied bands, have the same rotation invariants $\left(\left[X_{1}^{(2)}\right],\left[Y_{1}^{(2)}\right],\left[M_{1}^{(2)}\right]\right)=(-2,-2,-2)$. Then, at a disclination with Frank angle $\Omega=\pi$ and $\left[{\bf a}\right]^{(2)}=(0,0)$, both the obstructed atomic insulator and fragile TCI have fractional charge $\frac{1}{2}$. 
Therefore, one \emph{cannot} generically use fractional disclination charge as an indicator to distinguish fragile TCIs from obstructed atomic insulators as discussed in Ref \onlinecite{liu2018shift}. For some models it may be appropriate, but we offered a counter example above.

\subsection{Numerical simulation of fractional disclination charge}
\label{sec:numerics}
\begin{figure}[t!]
\centering
\includegraphics[width=\columnwidth]{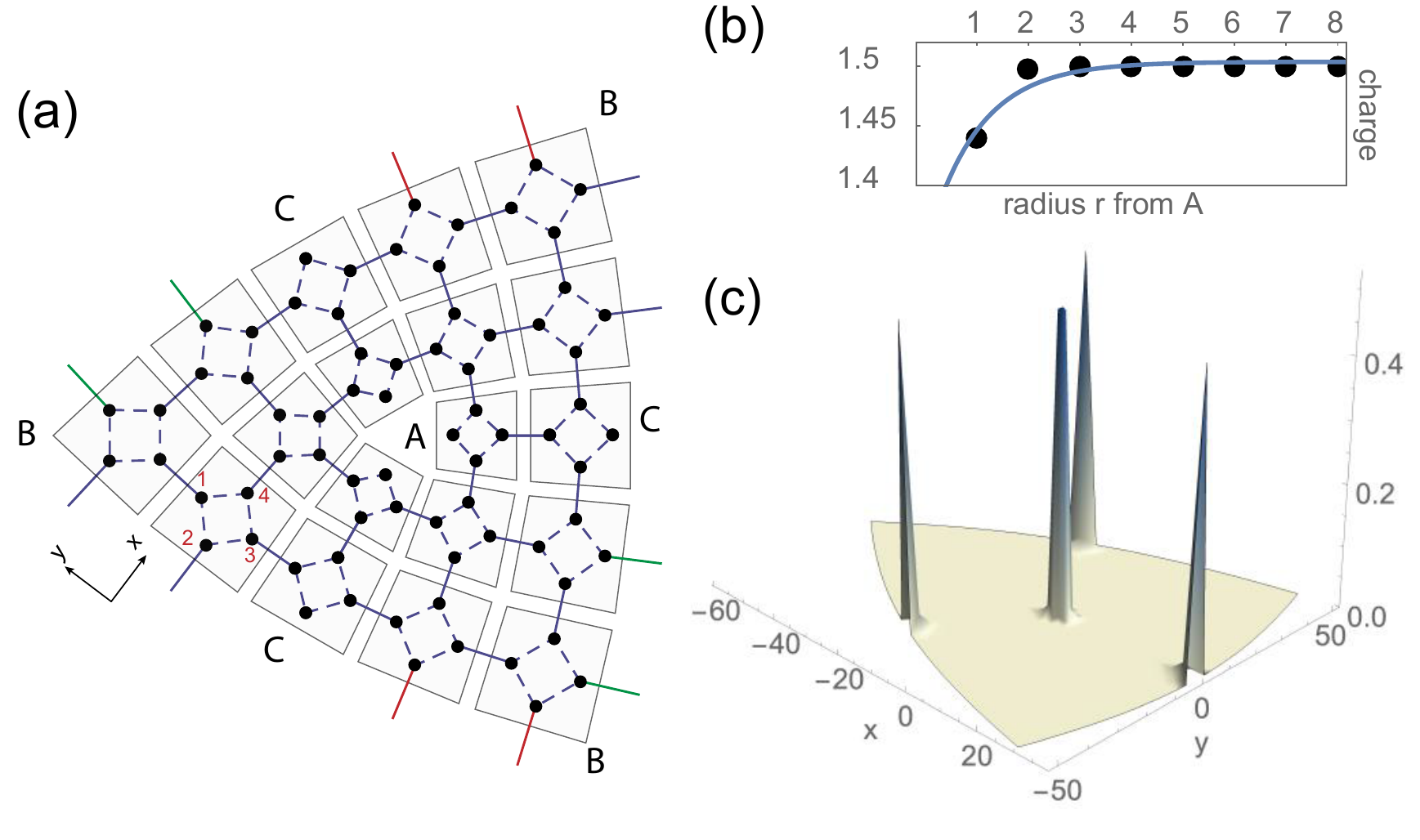}
\caption{(Color online) (a) Lattice configuration of the model in Eq.~\eqref{Eq:H_2c} on a periodic lattice with three disclinations. The gray squares and black dots represent the unit cells and the sublattice sites, respectively. Solid (Dashed) lines represent the nearest neighboring inter-cell (intra-cell) hopping terms. Periodic boundaries are identified by solid lines with the same color. There are three disclinations in this lattice, which are centered at $A, B$ and $C$ respectively. $A$ has $\Omega=-\frac{\pi}{2} [\bd{a}]^{(4)}=0$; $B$ has $\Omega=-\frac{\pi}{2},[\bd{a}]^{(4)}=1$; and $C$ has $\Omega=\pi,[\bd{a}]^{(2)}=(1,0)$. (b) The integral of charge density over a radius $r$ away from A. The curve is an exponential fit for the data points. The decay length is proportional to the correlation length, and thus proportional to the inverse of the energy gap. (c) Charge density distribution for the lattice in (a) with $31$ unit cells in each side.}
\label{fig:numericaH2c}
\end{figure}

In this section, to numerically verify our indices, %apply beyond the zero-correlation length (pictorial) limit 
we %numerically 
study the charge distribution for a spinless $C_4$ symmetric tight-binding model on a lattice with multiple disclinations. We verify that the indices we derived in Sec. \ref{sec:fracdisrotidx} correctly capture the fractional charge localized at the disclination cores. Figure~\ref{fig:numericaH2c}(a) shows the lattice configuration of the model with disclinations. It has four sites (black dots) per unit cell (the gray square) and nearest-neighbor intra-cell (dashed lines) and inter-cell (solid lines) hopping terms. In a defect-free lattice with translation symmetry, and with the basis of sublattice sites labeled in Fig.~\ref{fig:numericaH2c}(a), the Bloch Hamiltonian is,
\begin{align}
\label{Eq:H_2c}
H^{(4)}(\bd{k})=
\setlength{\arraycolsep}{0.1pt}
\begin{pmatrix}
0 & 0 & t_0+e^{i k_x} & 0 \\
0 & 0 & 0 & t_0+e^{i k_y} \\
t_0+e^{-i k_x}& 0 & 0 & 0 \\
0 & t_0+e^{-i k_y}& 0 & 0
\end{pmatrix},
\end{align}
where $t_0$ is the amplitude of the intra-cell hopping terms, and we have set the amplitude of the inter-cell hopping to $1$. 
The $C_4$ rotation operator is
\begin{align}
\label{rot_c4}
\hat{r}_4=\left(\begin{array}{cccc}
0 & 0 & 0 & 1\\
1 & 0 & 0 & 0\\
0 & 1 & 0 & 0\\
0 & 0 & 1 & 0\\
\end{array}\right),
\end{align}
which satisfies $\hat{r}_4^4=1$.

If $|t_0|\neq 1$ the bulk bands are gapped at half-filling and we choose to fill the lowest two bands.
When $t_0<1$, the model is in the topological phase. We summarize the eigenvalues of rotation operators at HSPs for the lowest two bands in Table~\ref{tab:Band Reps}. The resulting rotation invariants are $[X_1]=1, [M^{(4)}_1]=1,[M^{(4)}_3]=1$, indicating a nontrivial bulk polarization $\bd{P}=\frac{1}{2}({\bf{e}}_1+{\bf{e}}_2)$. Comparing the band representation with the band representations induced by different $\alpha_l$'s, we find the lowest two bands have the Wannier configuration $\{n_{c}^{1}=1\}$, corresponding to one Wannier orbital at each of the two points of the Wyckoff position $c$ [see Fig.~\ref{fig:WyckoffPos}(a)] per unit cell.
\begin{table}[t]
\renewcommand{\arraystretch}{1.5}
\caption{The band representation of model in Eq.~\eqref{Eq:H_2c} with eigenvalues of the $C_4$ rotation at $\bd{\Gamma}$ and $\bd{M}$ point and the $C_2$ rotation at $\bd{X}$ point. }
\label{tab:Band Reps}
    \centering
    \begin{tabular}{p{0.23\columnwidth}|p{0.23\columnwidth}p{0.23\columnwidth}p{0.23\columnwidth}p{0.23\columnwidth}}
    \hline\hline
    $C_4$ HSPs & ${\bf \Gamma}$ & ${\bf M}$ & ${\bf X}$ \\
    $H^{(4)}(\bd{k})$  & $i,-i $ & $-1,1$ & $-1,1$\\
    \hline\hline
    \end{tabular}
\end{table}

We can construct the lattice with multiple disclinations by connecting three rectangular patches [area circled by gray dashed line in Fig.~\ref{fig:numericaH2c}(a)] together with the inter-cell hopping terms. The unit cells at the boundary of each patch having hopping terms with the same color are connected. With this choice of boundary condition, the center of the lattice [$A$ in Fig.~\ref{fig:numericaH2c}(a)] forms a disclination with $\Omega=-\frac{\pi}{2},[\bd{a}]^{(4)}=0$ [see Fig.~\ref{fig:ConstructDis}(b)]. The translation part of the holonomy is $\bd{a}=-2\bd{e}_1$ corresponding to a normal vector $\bd{T}^{(n)}=-2\bd{e}_2$. As a result, the term related to the bulk polarization in Eq.~\eqref{eq:dischargek}, $\bd{T}^{(4)}\cdot\bd{P}^{(4)}$ is an integer and does not contribute to the fractional disclination charge. Therefore, only the rotation part of the holonomy contributes to the fractional disclination charge at $A$. Substituting the rotation invariants into Eq.~\eqref{eq:dischargek}, the fractional disclination charge is $\frac{1}{2}\mod 1$. The middle points $C$ at each of the three edges in Fig.~\ref{fig:numericaH2c}(a) form a disclination with
$\Omega=\pi,[\bd{a}]^{(2)}=(1,0)$ [see Fig.~\ref{fig:ConstructDis}(c), and the corner $B$ forms a disclination with $\Omega=-\frac{\pi}{2},[\bd{a}]^{(4)}=1$ [see Fig.~\ref{fig:ConstructDis}(d)]. In both cases, the translation part of the holonomy contributes to the fractional disclination charges. However, substituting the rotation invariants into Eq.~\eqref{eq:dischargek}, at disclination $B$, the fractional charges from the rotation part of the holonomy compensate the fractional charges from the translation holonomy, resulting an integer charge at $B$. While at $C$, the rotation part of the holonomy does not induce fractional charges, leading to a fractional disclination charge of $\frac{1}{2}$. 

In Fig.~\ref{fig:numericaH2c}(c), we show the numerical simulation of the charge density after subtracting the background charge of two electrons per unit cell in the bulk. The net electronic charge is exponentially localized at the disclination cores $A$ and $C$. In Fig.~\ref{fig:numericaH2c}(b), we show the integrated charge density over an area enclosed by a circle with a radius $r$ away from the disclination core $A$. It converges to the quantized value $\frac{1}{2}\mod 1$ exponentially. At $C$, the integrated charge density also exponentially converges to the quantized value $\frac{1}{2}\mod 1$. Finally, at disclination $B$ the integrated net charge densities are zero. In all cases, the numerical results are consistent with the results calculated from the indices Eq.~\eqref{eq:dischargek}.

Finally, let us point out a subtlety in calculating the disclination charge using the indices. The correspondence between the Wannier orbitals in real space and symmetry representations at HSPs in ${\bf k}-$space requires the same choice of unit cells. That means, we do not cut inside unit cells when taking off or inserting sectors into the lattice to construct the disclinations. 
For example, in the recent work of Ref \onlinecite{liu2018shift}, Liu \emph{et al}. considered the fractional disclination charge in a $C_6$ symmetric fragile TI. When constructing a disclination of $\Omega=-\frac{\pi}{3}$, one needs to choose the nearest $6$ sites on the same hexagon to form a $C_6$ symmetric unit cell. Under that enlarged choice of unit cell, we calculated the fractional disclination charge through the rotation invariants of the Hamiltonian with enlarged unit cells. The results are consistent with the numerical simulations in Ref.~\onlinecite{liu2018shift}. This clearly shows that
the physics that the indices capture is also applicable to fragile TCIs. 
Therefore, we have numerically verified that our indices are reliable for fractional disclination charge in both fragile TCIs and atomic insulators, no matter whether it is induced by the translation part or rotation part of the holonomy. We will go on to discuss extensions of our indices to include Chern numbers in Sec. \ref{Sec:formulawithch}, but first we will conclude this section with a discussion of the stability of the fractional charges in the presence of symmetry-preserving interactions. 
% 
%===========
\subsection{Robustness of the
Fractional charge in the presence of interactions}
\label{sec:interaction}
In this subsection, we will determine the robustness of the fractional disclination charges to the presence of symmetry-preserving electron-electron interactions. We will focus on the stability of fractional charges in interacting phases that are adiabatically connected to the non-interacting TCIs, and leave the properties of new phases that are only induced in the presence of interactions to future work. 
For our arguments we require that the allowed interaction terms in a Hamiltonian should preserve the lattice translation symmetry, $C_{n}$ rotation symmetry, and $U(1)$ charge conservation symmetry. %
What is more, we require the interactions to be local.
Due to the lattice translation symmetry, we can reduce the entire lattice to one unit cell and study the charge (corresponding to the $U(1)$ symmetry) and angular momentum (corresponding to $C_{n}$ symmetry) per unit cell. 

Let us consider a $C_{n}$ symmetric TCI which can be represented as $\sum_{\alpha,l}n_{\alpha}^{l}\alpha_{l}$.
The charge per unit cell is $Q\equiv \sum_{\alpha,l}M_{\alpha}n_{\alpha}^{l}$, where $M_{\alpha}$ is the multiplicity of Wyckoff position $\alpha$, and the angular momentum per unit cell is $L\equiv (\sum_{\alpha,l}lM_{\alpha}n_{\alpha}^{l}) \mod n$, where $l$ is the angular momentum of the Wannier orbitals localized at Wyckoff position $\alpha$(defined in Sec. \ref{Sec: fracdisWan}).
For example, in a $C_4$ symmetric TCI having Wannier representation $c_0$ which is induced from the trivial irrep of the site symmetry group $C_2$ at the Wyckoff position $c$, there are two s-orbitals at two equivalent Wyckoff positions $c$ (see Fig.~\ref{fig:WyckoffPos}) in each unit cell, and thus the total charge $Q=2$ and total angular momentum $L=M_{c}\times 0=0$. Similarly, in a TCI having Wannier representation $c_1$ which is induced from the non-trivial irrep of the site symmetry group $C_2$ at the Wyckoff position $c$, there are two p-orbitals at two equivalent Wyckoff positions $c$ (see Fig.~\ref{fig:WyckoffPos}), so the total charge is $Q=2$ and the total angular momentum is $L=M_{c}\times 1=2$.
Both the charge $Q$ and the angular momentum $L$ per unit cell should be conserved when we turn on interactions as we specified the interactions preserve $C_n$ and $U(1)$ symmetry.
Through a process where we  turn on interactions, perform a deformation, and then turn interaction off, we can deform a non-interacting TCI into another non-interacting TCI with different topology, i.e., different rotation invariants,   without closing the gap, which might cause a change in charge fractionalization at a disclination. In order to study whether the fractional charge at disclinations is robust when interactions are present, we first discuss the possible deformations in interacting systems\cite{else2019fragile}.
The conservation of charge and angular momentum imposes constraints on the allowed deformations that connect two insulators described by $\{n_{\alpha}^{l}\}$ and $\{(n_{\alpha}^{l})^{\prime}\}$ respectively: $Q^{\prime}=Q$ and $L^{\prime}=L$. With these two restrictions we have the following rules for allowed deformation:
\begin{enumerate}
\item{}At one specific Wyckoff position $\alpha$, a set of orbitals described by $\{n_{\alpha}^{l}\}$ is equivalent to another set of orbitals described by $\{(n_{\alpha}^{l})^{\prime}\}$  if  $Q^{\prime}=Q$ and $L^{\prime}=L$, i.e., they are connected by an adiabatic and continuous symmetry-preserving deformation. For example, in $C_{4}$ symmetric case, two filled $a_{1}$'s are equivalent to two filled $a_{3}$'s.
\item{} For a specific Wyckoff position $\alpha$, let the orbital configuration be described by $\{n_{\alpha}^{l}\}.$ Every complete set of all possible representations $l$ of $C_n$ located at $\alpha$  can be continuously deformed into a set of orbitals centered at another Wyckoff position $\beta$ through a continuously deformable general Wyckoff position. The new configuration $n_{\beta}^{l}$ will also contain (at least one) complete set of representations. For example, a $C_{4}$ symmetric TCI $a_{0}+a_{1}+a_{2}+a_{3}$ can be continuously deformed to a $C_4$ symmetric TCI $c_{0}+c_{1}$, both of which have charge $Q=4$ and angular momentum $L=2 \ {\rm mod} \ 4$.
\end{enumerate}
Note that the first rule is only true when there are interactions because if there is no interaction, the band gap, or equivalently, the gap between the many-body ground state and excited states must close during the deformation. Let us again take the $C_{4}$ symmetric case as an example. 
When there is no interaction, if we want to deform two
$a_{1}$'s into two $a_{3}$'s, the band gap must close and reopen because the occupied bands with eigenvalue $i$ must touch the conduction bands with eigenvalue $-i$ to exchange the eigenvalue of the rotation operator. This means the gap between the many-body ground state and excited states must close and reopen. When there are interactions, however, we can have symmetry preserving interaction terms like $c_{3a,1}^{\dag}c_{3a,2}^{\dag}c_{1a,1}c_{1a,2}$, where $c_{la,j}$ destroys an electron represented by $a_{l}$ and $j$ is the index for the internal degrees of freedom. This kind of interaction term can keep the gap between many-body ground state and excited states open during the deformation. What is more, under the first kind of deformations, since the interactions are local, the change of the configuration of Wannier orbitals happens locally, $i.e.$, the Wannier orbitals localized at a certain Wyckoff position are continuously deformed into Wannier orbitals localized at the same Wyckoff position though with different angular momentums. Thus, the localized Wannier states, or or their interacting analogs, always exist. As a result, if we integrate the charge density around the disclination core during the deformation where there are interactions, we can still get exponetially converged results, and thus the fractional discliantion charge is still well defined.

From these allowed deformations we can see that the fractional charge is stable against interactions in the sense that all adiabatic deformations that are not allowed %cause phase transition
in the non-interacting cases but are allowed in the cases with symmetric interactions will not change the fractional disclination charge.
% in the sense that even though the deformations can deform a TCI $A$ into a TCI $B$ which would be topologically distinct from $A$ in the free fermion case, the fractional disclination charge will not change}. 
The first kind of deformation \emph{cannot} change $n_{\alpha}=\sum_{l}n_{\alpha}^{l}$ for any $\alpha$, so the fractional charge will not change after the deformation according to Eq. \eqref{eq:discharge}. The second kind of deformation always changes  $M_{\alpha}n_{\alpha}$ by multiples of $n$ for a $C_{n}$ symmetric TCI. Thus, according to Eq. \eqref{eq:discharge}, the fractional charge will not change after the second deformation either.
If we further impose TRS with $T^{2}=-1$, the $n_{\alpha}=\sum_{l}n_{\alpha}^{l}$ still remains the same after the first kind of deformation and at each Wcykoff position $\alpha$, we must have Wannier orbitals of all angular momentums $l=\frac{1}{2},\frac{1}{2}...,m_\alpha-\frac{1}{2}$ and their Kramers' partners to form a complete set of representations of $C_n$, which is required for the second deformation. The change of $M_{\alpha}n_{\alpha}$ is therefore always multiples of $2n$ for a $C_{n}$ symmetric TCI after the second kind of deformation. Thus, according to \eqref{eq:discharge}, we can conclude that after the two allowed deformations, the fractional charge still remains the same, and furthermore, the oddness or evenness of the integer charge also does not change when imposing TRS with $T^{2}=-1$(Note that as discussed in Sec \ref{sec:fracdisrotidx}, the topological charge should be defined modulo $2$ when there is TRS with $T^{2}=-1$). In conclusion, the fractional disclination charge is robust in the presence of interactions. When there is TRS with $T^{2}=-1$, the oddness or evenness of the integer charge is also robust in the presence of interactions.

%===========
\section{Fractional Disclination Charge for TCIs with Chern number}
\label{Sec:formulawithch}
So far we have investigated the fractional disclination charges for two dimensional $C_n$ symmetric TCIs that admit a (generalized) Wannier representation. In this section, we will use an algebraic method to generalize the indices for fractional disclination charge to TCIs with a non-zero Chern number. In this case  one cannot construct spatially localized Wannier functions from the Bloch states of the occupied bands, and therefore a (generalized) Wannier picture is not applicable \cite{brouder2007exponential,soluyanov2011wannier}. However, it is known that the Chern number can survive in the presence of interactions\cite{niu1985} so we expect the contributions due to the Chern number to remain when interactions are turned on (at least when interactions preserve rotation symmetries). 

Each class of 2D $C_n$ symmetric TCIs can be characterized by a set of topological invariants, $\chi^{(n)}=\{[{\bf \Pi^{(n)}}],ch\}$, where $[{\bf \Pi}^{(n)}]$ is an irreducible set of rotation invariants, and $ch$ is the (first) Chern number. Consider two Hamiltonians $H_1$ and $ H_2$ with the same symmetries, but belonging to different classes with the topological invariants $\chi^{(n)}_1, \chi^{(n)}_2,$ and fractional disclination charges $Q_1, Q_2$. If we stack them together, the combined TCI can be described by a new Hamiltonian $H=H_1\oplus H_2$ with topological invariants $\chi^{(n)} = \chi^{(n)}_1+\chi^{(n)}_2$, where the Chern numbers add integers, and the rotation invariants add up according to their cyclic group structure. One can turn on symmetry preserving coupling terms between $H_1$ and $H_2$, and $\chi^{(n)}$ remains the same as long as the bulk gap is not closed. 

When stacking the two Hamiltonians together, the fractional disclination charge in the combined insulator is $Q=Q_1+Q_2,$ and it is stable to the addition of symmetry preserving coupling terms that do not close the band gap. Due to the same additive algebraic structure of the fractional disclination charge $Q,$ and the topological invariants $\chi^{(n)}$, the index of fractional disclination charge $Q(\chi^{(n)})$ is a linear function of $\{[{\bf \Pi}^{(n)}],ch\}$ for a given disclination type. Therefore, one can use the algebraic method in Refs.  \onlinecite{PhysRevLett.111.047006,benalcazar2014classification} to derive the index $Q(\chi^{(n)})$ from a set of primitive Hamiltonian generator models with linearly independent invariants for each $C_n$ symmetry. The disclination charge indices $Q_{dis}^{(n)}$ that we derived in Sec.~\ref{sec:fracdisrotidx} using the real-space Wannier picture and band representation theory, correspond to an index $Q(\chi^{(n)})$ derived from generators with $ch=0$. In order to generalize our indices to TCIs with non-zero Chern number, for a given $C_n$ symmetry and disclination type, we can set the disclination charge as
\begin{align}
Q_{dis,ch}^{(n)} = Q_{dis}^{(n)}([{\bf \Pi}^{(n)}]) +\beta ch \mod 1
\end{align}
where we use $Q_{dis,ch}^{(n)}$ to represent the generalized index of fractional disclination charge including Chern number contributions to distinguish it from the index $Q_{dis}^{(n)}$ in Sec.~\ref{sec:fracdisrotidx}. The parameter $\beta$ generically depends on the disclination type $(n_\Omega, {\bf a})$, and the problem of generalizing the fractional charge indices boils down to determining $\beta.$ 

To find $\beta$ we can proceed as follows. Consider a lattice having two disclinations $(\Omega_1,{\bf a}_1)$ and $(\Omega_2,{\bf a}_2)$, each of which binds a fractional charge $Q_{dis,ch}(\Omega_i,{\bf a}_i), i =1,2$. The total charge localized at the region covering the two disclinations is
\begin{align}
Q=&Q_{dis,ch}(\Omega_i,{\bf a}_i,ch)+Q_{dis,ch}(\Omega_2,{\bf a}_2, ch)\nonumber\\
=&Q_{dis,ch}(\Omega_1+\Omega_2,{\bf a}_1+\hat{r}(\Omega_1){\bf a}_2,ch).
\end{align}
Therefore, in order to be compatible with the addition rule of disclinations, $\beta$ must be proportional to $\frac{\Omega}{2\pi}$,
\begin{align}
Q_{dis,ch}^{(n)}=Q_{dis}^{(n)}+\gamma \frac{\Omega}{2\pi} ch.
\end{align}
To solve for $\gamma$ in each $Q_{dis,ch}^{(n)}$, we numerically calculated the fractional charge localized at disclinations for a $C_{n}$-symmetric insulator with $ch=1$. 
We summarize the results in the Table~\ref{tab:ch}. The details of the Hamiltonian of each Chern insulator, and corresponding charge density distribution are shown in Appendix \ref{sec: numchern}.
\begin{table}[ht]
\centering
\caption{Fractional charge modulo $1$ at disclination for $C_n$ symmetric insulator with $ch=1$.} 
\label{tab:ch}
\renewcommand{\arraystretch}{1.5}
\begin{tabular}{p{0.38\columnwidth}p{0.52\columnwidth}p{0.08\columnwidth}}
\hline\hline
Disclination type & Topological invariants & $Q_{dis,ch}$\\
\hline
$\Omega=\pi,[{\bf a}]^{(2)}=(1,0)$ & $[X^{(2)}_1]=[Y^{(2)}_1]=0, [M^{(2)}_1]=1$ & $\frac{1}{2}$\\
$\Omega=-\frac{\pi}{2},[{\bf a}]^{(4)}=0 $ & $[X^{(2)}_1]=[M^{(4)}_3]=0, [M^{(4)}_1]=1$ &  $0$\\
$\Omega=-\frac{2\pi}{3},[{\bf a}]^{(3)}=0 $ & $[K^{(3)}_2]=[K^{'(3)}_2]=-1$ &  $\frac{1}{2}$\\
$\Omega=-\frac{\pi}{3},[{\bf a}]^{(6)}\equiv 0 $ & $[K^{(3)}_2]=[M^{(2)}_1]=-1$ &  $\frac{1}{4}$\\
\hline\hline
\end{tabular}
\end{table}

Having established the numerical results in Table \ref{tab:ch}, we can solve for $\gamma$ and write the generalized indices:
\begin{equation}
\label{eq:dischargekch}
\begin{array}{l}
\begin{aligned}
Q_{dis,ch}^{(2)}=&\frac{\Omega}{2\pi}\left(\frac{1}{2}\left[X^{(2)}_{1}\right]+\frac{1}{2}\left[Y^{(2)}_{1}\right]+\frac{1}{2}\left[M^{(2)}_{1}\right]+\frac{1}{2}ch\right)
\\
&+\mathbf{T}^{(2)}\cdot\mathbf{P}^{(2)} \ \rm{mod} \ 1,
\end{aligned}
\\
\\
\begin{aligned}
Q_{dis,ch}^{(3)}=&\frac{\Omega}{2\pi}\left(\left[K^{(3)}_{2}\right]+\left[K^{\prime(3)}_{2}\right]+\frac{1}{2}ch\right)
\\
&+\mathbf{T}^{(3)}\cdot\mathbf{P}^{(3)} \ \rm{mod} \ 1,
\end{aligned}
\\
\\
\begin{aligned}
Q_{dis,ch}^{(4)}=&\frac{\Omega}{2\pi}\left(\left[X^{(2)}_{1}\right]+\frac{3}{2}\left[M_{3}^{(4)}\right]-\frac{1}{2}\left[M_{1}^{(4)}\right]+\frac{1}{2}ch\right)
\\
&+\mathbf{T}^{(4)}\cdot\mathbf{P}^{(4)}\ \rm{mod} \ 1,
\end{aligned}
\\
\\
\begin{aligned}
Q_{dis,ch}^{(6)}=\frac{\Omega}{2\pi}\left(2\left[K_{2}^{(3)}\right]+\frac{3}{2}\left[M^{(2)}_{1}\right]+2ch\right) \ \rm{mod} \ 1.
\end{aligned}
\end{array}
\end{equation}
To confirm these formulae we also numerically calculated the fractional charge at different types of disclinations for each $C_n$ symmetric insulator with different Chern numbers. All the numerical results are consistent with our indices for fractional disclination charge in Eq. \eqref{eq:dischargekch}(see Appendix \ref{sec: numchern} for more details).

What we find here is that disclinations, which are sources of geometric curvature, act as an effective flux. In the presence of a Chern number, and/or non-trivial rotation invariants, this flux traps charge. A similar effect is known in the quantum Hall effect literature where the $U(1)$ electromagnetic field can couple, in the presence of \emph{continuous} rotation symmetry, to the geometric spin-connection field with the Wen-Zee term\cite{wen1992shift,abanov2014electromagnetic,biswas2016,you2018higher,han2019generalized}
\begin{equation}
\mathcal{L}_{WZ}=\frac{ch}{2\pi}\omega\wedge dA.
\end{equation}
We can compare our results to the charge response of a quantum Hall insulator with a Chern number $ch$ from which we find
\begin{equation}
\label{eq: WZq}
\begin{aligned}
Q_{dis}^{WZ}=\frac{ch}{2\pi}\int_{\mathcal{S}}(\partial_{x}\omega_{y}-\partial_{y}\omega_{x})=\frac{ch}{2\pi}\oint_{\mathcal{C}}{\boldsymbol{\omega}}\cdot d{\boldsymbol{l}}
=\frac{\Omega}{2\pi}ch.
\end{aligned}
\end{equation}

For the $C_{6}$ symmetric case, according to Eq.\eqref{eq: Chern_rot}, we can substitute $2ch=ch+3\left[M_{2}^{(2)}\right]+4\left[K_{2}^{(3)}\right]+2\left[K_{3}^{(3)}\right]+6N$ into Eq. \eqref{eq:dischargekch} and find
\begin{equation}\label{eq:c6chargecompare}
Q_{dis,ch}^{(6)}=\frac{\Omega}{2\pi}\left(2\left[K_{3}^{(3)}\right]-\frac{3}{2}\left[M^{(2)}_{1}\right]+ch\right) \ \rm{mod} \ 1.
\end{equation}
Note that we include the $6N$ (where $N$ is an undetermined integer) is because that the relationship between the Chern number and rotation invariants in Eq. \eqref{eq: Chern_rot} is only correct modulo 6. It is clear that the $6N$ term will not affect the fractional charge in Eq.~\eqref{eq:c6chargecompare} modulo 1.
We can conclude from this result that the coefficient in front of $ch$ in Eq.~\eqref{eq:c6chargecompare} is consistent with that in the Wen-Zee term in Eq. \eqref{eq: WZq}. However, if we try to repeat the process for the $C_{2}$, $C_{3}$ and $C_{4}$ symmetric cases we find results that do not agree.  Let us take $C_{2}$ symmetric case as an example. If we want to generate a term $\frac{\Omega}{2\pi}ch$ in $Q^{(2)}_{dis,ch}$ in Eq.\eqref{eq:dischargekch} that will match the Wen-Zee result, we can use Eq.\eqref{eq: Chern_rot} to rewrite $\frac{1}{2}ch$ as $\frac{1}{2}ch=ch-\frac{1}{2}\left(\left[X_{1}^{(2)}\right]+\left[Y_{1}^{(2)}\right]-\left[M^{(2)}_{1}\right]\right)-N^{\prime}$ and substitute it into the $C_2$ index in Eq. \eqref{eq:dischargekch}, where $N^{\prime}$ is an undetermined integer. Unlike the $C_{6}$ symmetric case, we cannot drop $N^{\prime}$ here because when $\Omega=\pi$, we will have a term $\frac{1}{2}N^{\prime}$ in $Q^{(2)}_{dis}$ that contributes to the fractional charge. For this reason, we cannot make the coefficients of the $ch$ consistent with the Wen-Zee result unless we introduce an undetermined term into the indices $Q_{dis, ch}^{(n)}$. Instead, we generically have coefficients of $ch$ that are \emph{half} of that in Eq.\eqref{eq: WZq}, \emph{i.e}, $\frac{\Omega}{4\pi}$ in $C_{2}$, $C_{3}$ and $C_{4}$ symmetric cases. It is unclear what final conclusions to draw about the connection between our invariants for discrete rotation symmetries and the Wen-Zee term that applies for continuous rotation symmetry. This is an exciting open problem for future work.

\section{Discussion and Conclusions}
\label{conclusionandcomment}
In conclusion, we have derived general indices for fractional disclination charge in two-dimensional $C_{n}$-symmetric TCIs. For TCIs that admit a (generalized) Wannier representation, including atomic limit insulators and fragile TCIs, the method we use clearly
illustrates the origin of charge fractionalization and the robustness of fractional disclination charge in the presence of interactions. For TCIs with non-zero Chern number, on the other hand, the lack of a pictorial representation makes the fractionalization of charge at disclinations far from evident. However, in this case, we can build the topological indices (Eq. \eqref{eq:dischargekch}) using an algebraic method. It is well known that an external flux piercing the plane in a Chern insulator leads to a charge response proportional to the product of Chern number and the flux. From our indices in Eq. \eqref{eq:dischargekch}, we can see that the Chern number couples to the Frank angle $\Omega$, suggesting that the disclinations can induce an effective flux, which is responsible for the appearance of fractional charges.

However, the exact correspondence between the effective flux and defects is an open question. Knowing the answer of this question will help identify fractional charges localized at defects with fewer symmetry requirements. For example, retaining translation symmetry in the bulk far from the defects may not be necessary for charge fractionalization at disclinations in Chern insulators, as it seemingly is for atomic limit insulators and fragile TCIs\cite{he2018chern}. That is,  the fractional charges at defects in a Chern insulator might be more robust against disorder. A related open question is the origin of the rotational invariant terms in Eq. \eqref{eq:dischargekch} in systems that do not admit (generalized) Wannier representation. These facts imply that there is richer physics besides the Wen-Zee response underlying charge fractionalization in Chern insulators with discrete rotation symmetries, and it deserves further investigation.
\acknowledgments{We thank Mao Lin and Ailei He for useful discussions. We thank Saavanth Velury for his feedback and help with manuscript preparation. TL, PZ, and TLH thank the US National Science Foundation (NSF) grants EFMA-1627184 (EFRI),
DMR-1351895, and the MRSEC program under NSF Award
Number DMR-1720633 (SuperSEED) for support. WAB thanks the support of the Eberly Postdoctoral Fellowship at the Pennsylvania State University.}
\bibliography{ref-1}

%==================================================
\appendix
\section{Graphs of different types of disclinations}
\label{sec: disclinations}
\begin{figure}[h!]
\centering
\includegraphics[width=1\columnwidth]{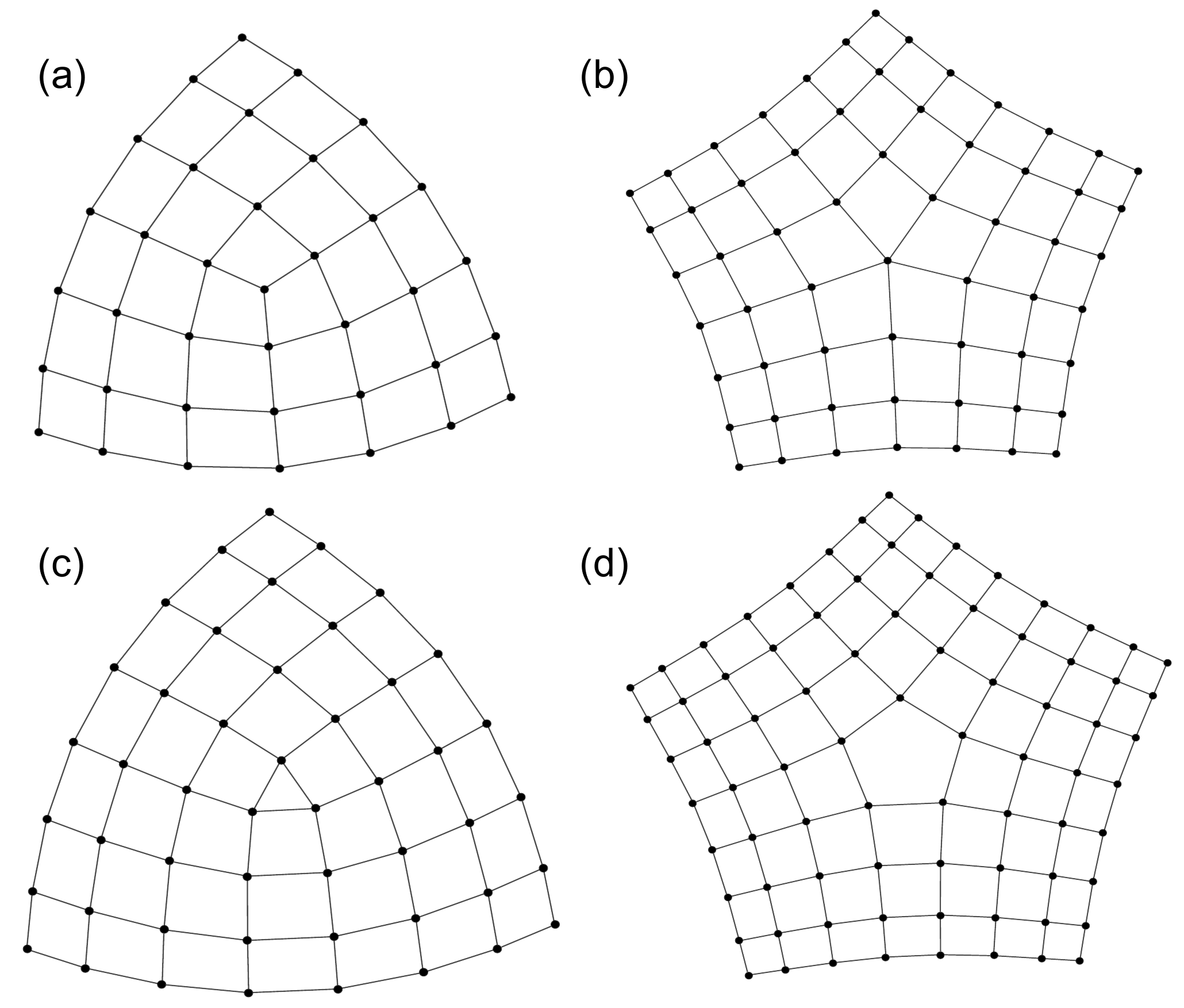}
\caption{Disclinations in $C_{4}$ symmetric lattices. (a) and (c) have Frank angle $\Omega=-90^{\circ}$; (b) and (d) have Frank angle $\Omega=+90^{\circ}$. (a) and (b) have $\left[{\bf a}\right]^{(4)}=0$; (c) and (d) have $\left[{\bf a}\right]^{(4)}=1$.}
\label{fig:c4dis}
\end{figure}

\begin{figure}[t!]
\centering
\includegraphics[width=1\columnwidth]{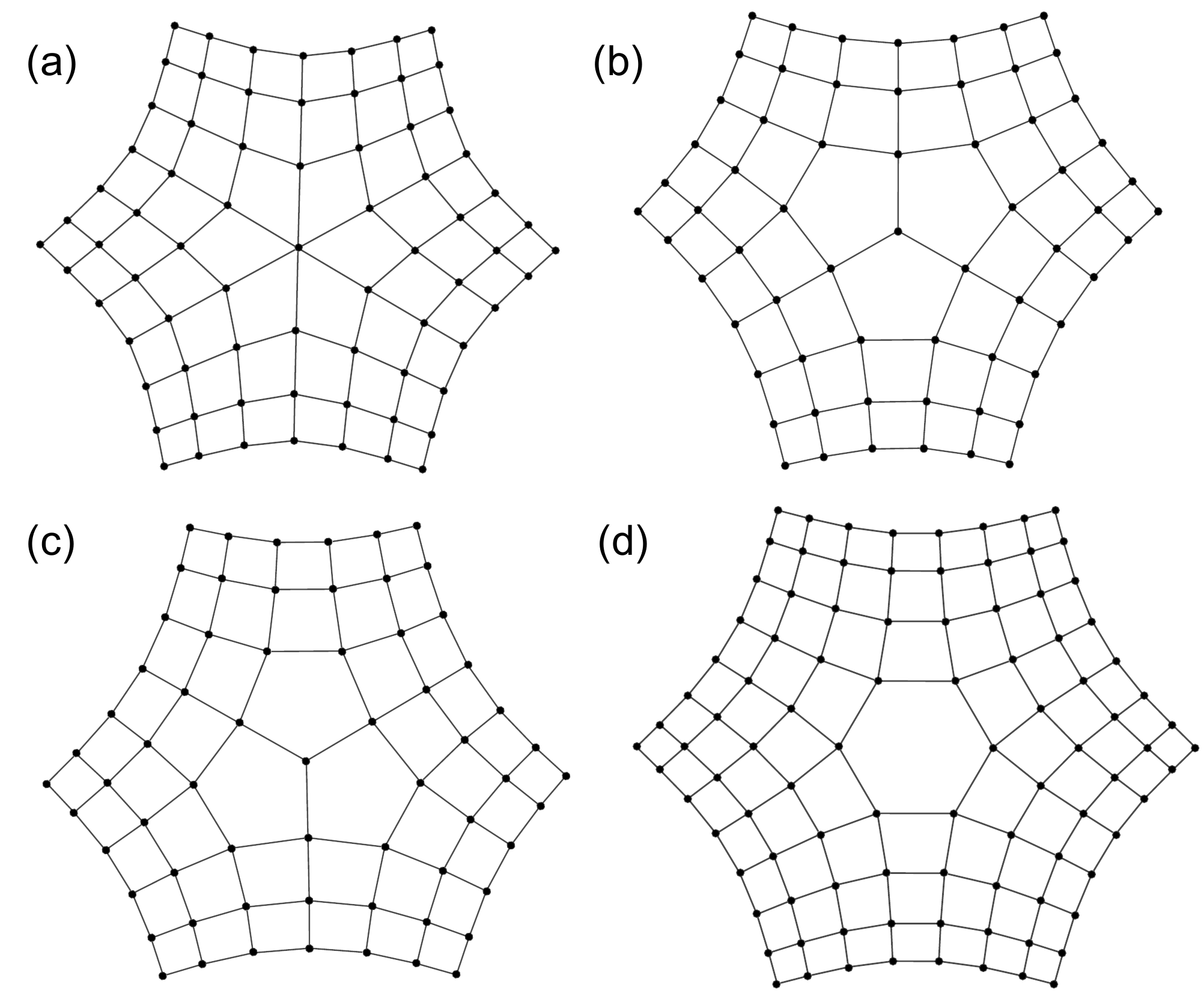}
\caption{Disclinations in $C_{2}$ symmetric lattices. $\Omega=180^{\circ}$ diclinations with different translation holonomy: (a) $\left[{\bf a}\right]^{(2)}=(0,0)$ (b) $\left[{\bf a}\right]^{(2)}=(0,1)$, (c) $\left[{\bf a}\right]^{(2)}=(1,0)$ and (d) $\left[{\bf a}\right]^{(2)}=(1,1)$.}
\label{fig:c2dis}
\end{figure}

\begin{figure}[t!]
\centering
\includegraphics[width=1 \columnwidth]{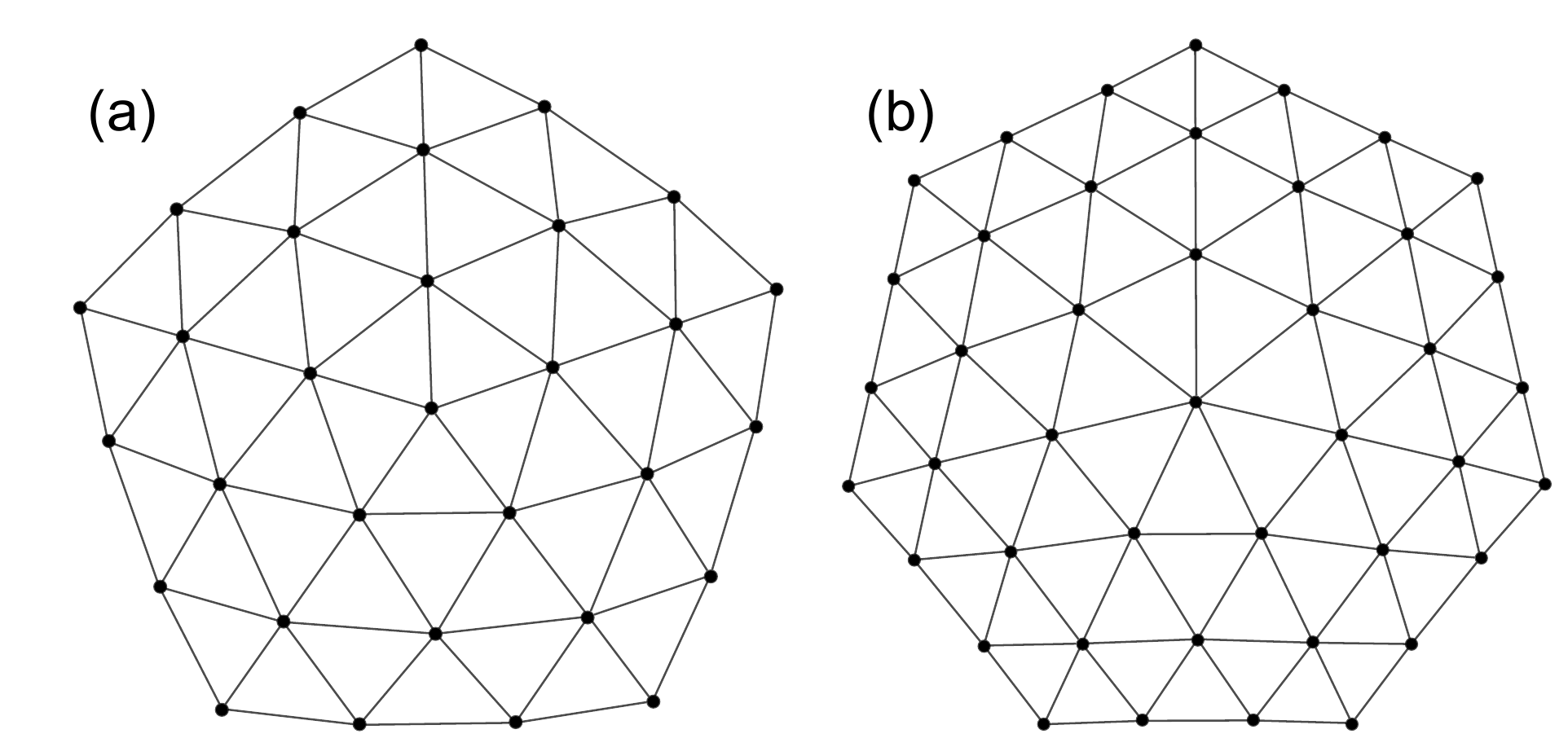}
\caption{Disclinations in $C_{6}$ symmetric lattices. (a) has Frank angle $\Omega=-60^{\circ}$; (b) has Frank angle $\Omega=+60^{\circ}$. }
\label{fig:c6dis}
\end{figure}

\begin{figure}[h!]
\centering
\includegraphics[width=1\columnwidth]{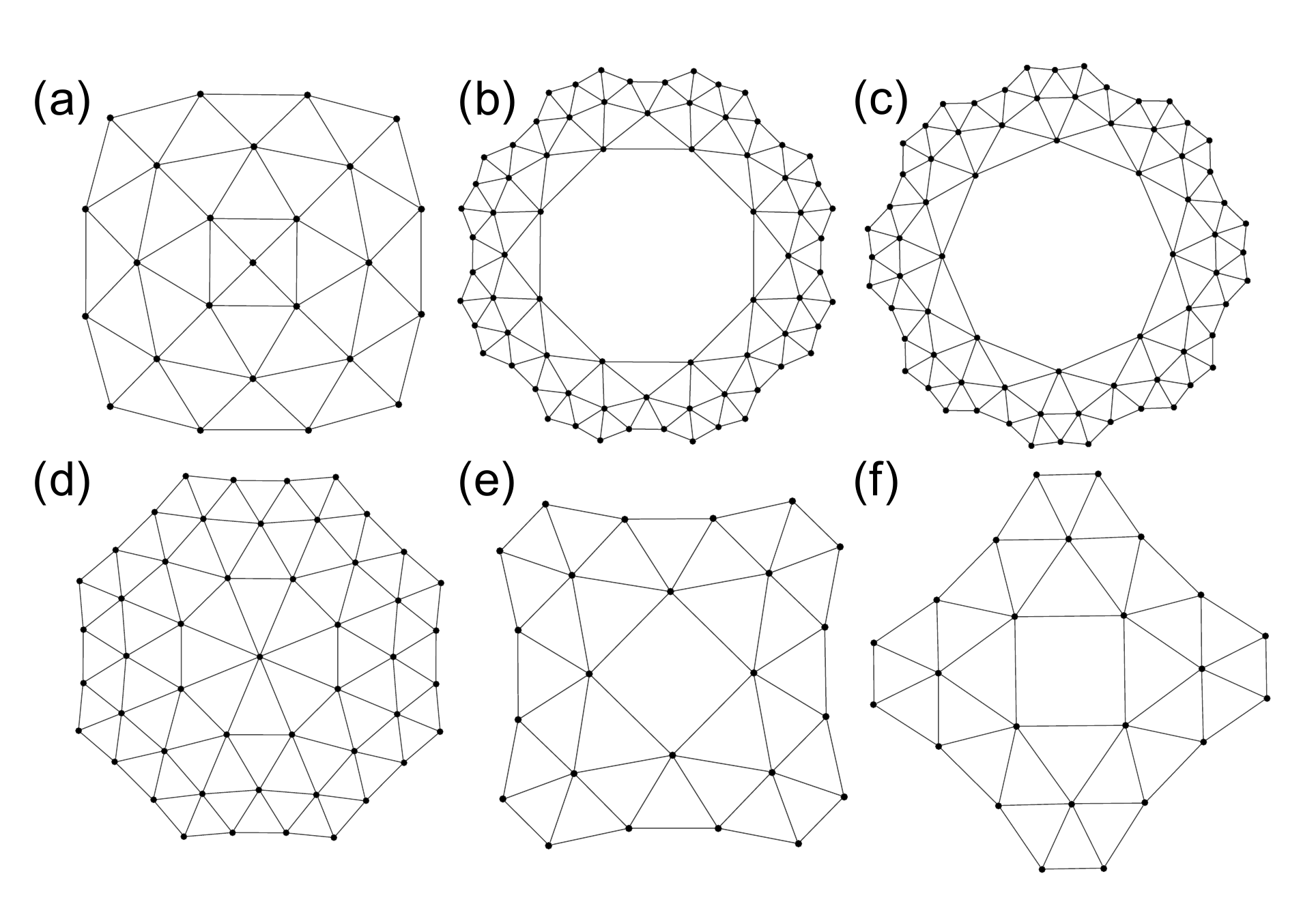}
\caption{Disclinations in $C_{3}$ symmetric lattices. (a,b) and (c) have Frank angle $\Omega=-120^{\circ}$; (d,e) and (f) have Frank angle $\Omega=+120^{\circ}$. (a,d) have $\left[{\bf a}\right]^{(3)}=0$; (b,e) have $\left[{\bf a}\right]^{(3)}=+1$; (c,f) have $\left[{\bf a}\right]^{(3)}=-1$. }
\label{fig:c3dis}
\end{figure}

In this section, we list lattices with disclinations of different holonomy $(\Omega,\left[{\bf a}\right]^{(n)})$. The Frank angles of disclinations in $C_n$ symmetric lattices are multiples of $\frac{2\pi}{n}$. For $C_2$ and $C_4$ symmetric TCIs, we use square lattices. Disclinations with Frank angle $\Omega=\pm\frac{\pi}{2}$ are shown in Fig.~\ref{fig:c4dis} and disclinations with Frank angle $\Omega=\pi$ are shown in Fig.~\ref{fig:c2dis}. For $C_3$ and $C_6$ symmetric TCIs, we use triangular lattice.  Disclinations with Frank angle $\Omega=\pm\frac{\pi}{3}$ are shown in Fig.~\ref{fig:c6dis} and disclinations with Frank angle $\Omega=\pm\frac{2\pi}{3}$ are shown in Fig.~\ref{fig:c3dis}. 
\section{Relation between the number of Wannier orbitals and rotation indices }
\label{sec: BandRep}
\subsection{Band representations}
The theory of band representations was originally introduced in Refs. \onlinecite{ZakBandRep1980,ZakBandRep1981}, and later generalized in Refs. \onlinecite{bradlyn2017topological,po2017symmetry,JenniferBuilding2018,bradlyn2019disconnected,kruthoff2017topological}. Here, we list some important conclusions from Ref \onlinecite{JenniferBuilding2018} which are useful to our discussion. 
Generally, the crystalline symmetry of the entire lattice can be described by a space group $\mathbf{G}$. There are high symmetry points ${\bf q}$ in real space that are invariant under a subgroup of ${\bf G}$, which is called the site symmetry group ${\bf G_q}$ for the point ${\bf q}$. 
Given a high symmetry point ${\bf q}_1$ in real space, we find $g_{\mu}\in \mathbf{G}$ satisfying $g_{\mu}\notin \mathbf{G}_{\mathbf{q}_1}$ to map ${\bf q}_1$ to other points in the unit cell of ${\bf q}_1$,  $g_{\mu}\mathbf{q}_{1}=\mathbf{q}_{\mu}$ for $\mu=2,3,\ldots M_\alpha$. These $M_\alpha$ points belong to the same Wyckoff position $\alpha$ with the multiplicity $M_\alpha$.
We can induce a representation of the space group $\mathbf{G}$ from an irreducible representation of the site symmetry group $\mathbf{G}_{\mathbf{q}_{1}}\in \mathbf{G}$ at the point $\mathbf{q}_{1}$ in the Wyckoff position $\alpha$. We denote the irreducible representation of $\mathbf{G}_{\mathbf{q}_{1}}$ by $\rho$ and denote the induced representation of $\mathbf{G}$ from $\rho$ by $\rho_{\mathbf{G}}$. For all $h \in \mathbf{G}$, we have
\begin{equation}
\label{eq:multirule}
hg_{\mu}=\{E|\mathbf{t}_{\nu\mu}\}g_{\nu}g,
\end{equation}
where $g\in \mathbf{G}_{\mathbf{q}}$, and ${\bf t}$ is a translation by a lattice vector, $\mathbf{t}_{\nu\mu}=h\mathbf{q}_{\mu}-\mathbf{q}_{\nu}$. 
Then, the induced representation at each ${\bf k}$, $\rho_{\mathbf{G}}^{\mathbf{k}}(h)$, is given by
\begin{equation}
\label{eq:bandrep}
\rho_{G}^{\bold{k}}(h)_{j\nu,i\mu}=e^{-i (R\mathbf{k})\cdot \bold{t}_{\nu\mu}}\rho_{ji}(g_{\nu}^{-1}\{E|-\mathbf{t}_{\nu\mu}\}hg_{\mu}),
\end{equation}
where $i,j=1,2,\cdots,n_{\mathbf{q}}$, and $n_{\mathbf{q}}$ is the dimension of the irreducible representation $\rho$ of the site symmetry group $G_{\mathbf{q}}$.

\begin{figure}[t]
\centering
\includegraphics[width=\columnwidth]{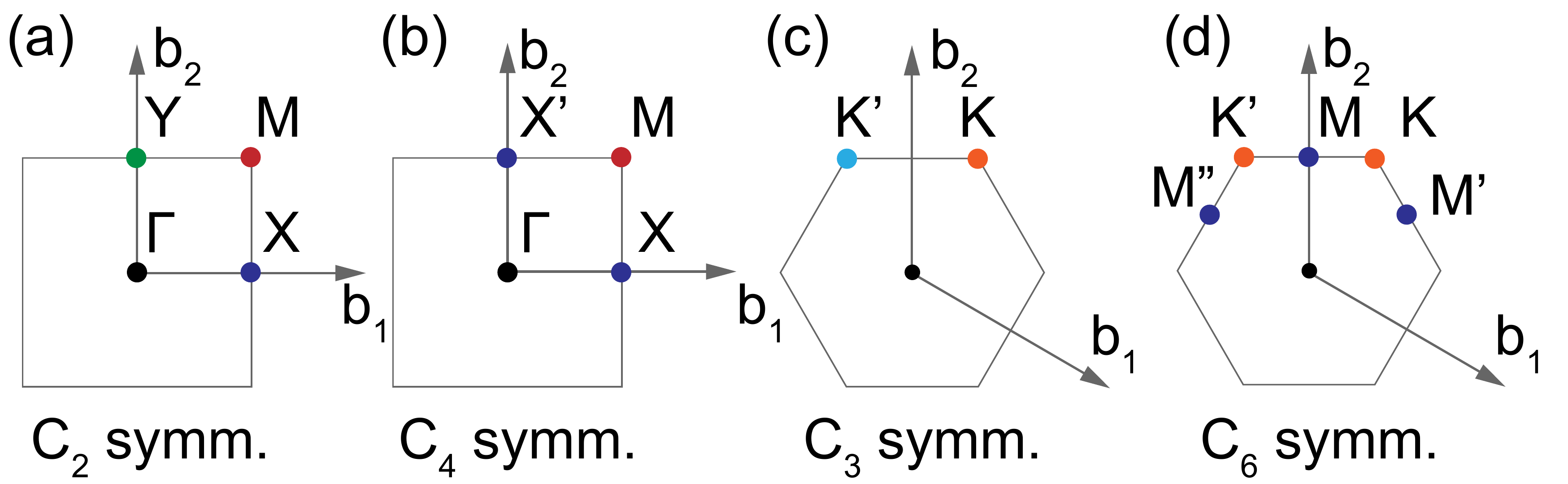}
\caption{(Color online) The Brillouin zone, reciprocal lattice vectors ${\bf b}_1,{\bf b}_2$ and HSPs for (a) $C_2$, (b) $C_4$ symmetric TCIs with $\bd{b}_1=\hat{x},\bd{b}_2=\hat{y}$; (c) $C_3$ and (d) $C_6$ symmetric TCIs with $\bd{b}_1=2\pi\left(\hat{x}-\frac{1}{\sqrt{3}}\hat{y}\right),\bd{b}_2=\frac{4\pi}{\sqrt{3}}\hat{y}$.}
\label{fig:BZ}
\end{figure}

For $C_n$ symmetric TCIs, $h=\{C_{n}|(0,0)\}$, we can use Eq.~\eqref{eq:bandrep} to calculate the rotation eigenvalues of an induced band representation $\rho_{G}^{\bold{k}}(\{C_{n}|(0,0)\})$ from the irreducible representation of the site symmetry group of a maximal Wyckoff position at all HSPs in ${\bf k}$ space (see Fig.\ref{fig:WyckoffPos} for all maximal Wyckoff positions).
Fig.~\ref{fig:BZ} shows the BZ and HSPs for $C_2,C_3,C_4$ and $C_6$ TCIs.
In the following, we will use Eq.~\eqref{eq:bandrep} to derive the relation between the number of Wannier orbitals at a maximal Wykcoff position and rotation invariants at HSPs in ${\bf k}-$space for each $C_n$ symmetry. 

\subsection{Twofold symmetry}
\label{subsec:twofold}
As shown in Fig.~\ref{fig:WyckoffPos}, the four maximal Wyckoff positions of the $C_{2}$ symmetric lattice are labelled by a, b, c, and d. We choose $\mathbf{e}_{1}=(1,0)$ and $\mathbf{e}_{2}=(0,1)$ to be the lattice vectors. The multiplicities of these four Wyckoff positions are all $1$. Thus, we can ignore the indices $\mu$ and $\nu$ in Eq.~\eqref{eq:bandrep}. In order to induce the band representation of $\mathbf{G}$, which is generated by the $C_{2}$ rotation and lattice translations, from the irreducible representation $\rho$ of $\mathbf{G}_{a}=\left\{\{E|(0,0)\},\{C_{2}|(0,0)\}\right\}$, we substitute $h=\{C_{2}|(0,0)\}$ into Eq.\eqref{eq:bandrep} and find
\begin{equation}
\label{eq:bandrepc2_1}
\rho_{G}^{\bold{k}}(\{C_{2}|(0,0)\})=\rho(\{C_{2}|(0,0)\}).
\end{equation}
Since $C_{n}$ group is Abelian, the indices $i$ and $j$ are also unnecessary, i.e., $\rho(g)$ for $\forall g\in \mathbf{G}_{a}$ is just a number. 
For the Wyckoff positions $b, c,$ and $d$, with site symmetry groups $\mathbf{G}_{b}=\left\{\{E|(0,0)\},\{C_{2}|(1,1)\}\right\}, \mathbf{G}_{c}=\left\{\{E|(0,0)\},\{C_{2}|(1,0)\}\right\}$ and $ \mathbf{G}_{d}=\left\{\{E|(0,0)\},\{C_{2}|(0,1)\}\right\}$, respectively, the induced representations are,
\begin{equation}
\label{eq:bandrepc2_2}
\begin{array}{l}
\rho_{G}^{\bold{k}}(\{C_{2}|(0,0)\})=e^{-i (R_{2}\bold{k})\cdot (-1,0)}\rho(\{C_{2}|(1,0)\}),
\\
\rho_{G}^{\bold{k}}(\{C_{2}|(0,0)\})=e^{-i (R_{2}\bold{k})\cdot (0,-1)}\rho(\{C_{2}|(0,1)\}),
\\
\rho_{G}^{\bold{k}}(\{C_{2}|(0,0)\})=e^{-i (R_{2}\bold{k})\cdot (-1,-1)}\rho(\{C_{2}|(1,1)\}).
\end{array}
\end{equation}

From these band representations, we can read off the eigenvalues for the 2-fold rotation operator ($\{C_{2}|(0,0)\}$) at all HSPs. The results are listed in Table~\ref{Tab:Wy_C2}, 
\begin{table}[h!]
\centering
\begin{tabular}{p{0.3\columnwidth}|p{0.15\columnwidth}p{0.15\columnwidth}p{0.15\columnwidth}p{0.15\columnwidth}}
\hline\hline
Wyckoff positions & $\bd{\Gamma}$ & $\bd{M}$ & $\bd{X}$ & $\bd{Y}$\\
\hline
$a_l$ & $e^{il\pi}$ & $e^{il\pi}$
      & $e^{il\pi}$ & $e^{il\pi}$ \\
$b_l$ & $e^{il\pi}$ & $e^{il\pi}$
      & $e^{i(l+1)\pi}$ & $e^{i(l+1)\pi}$ \\
$c_l$ & $e^{il\pi}$ & $e^{i(l+1)\pi}$
      & $e^{i(l+1)\pi}$ & $e^{il\pi}$ \\
$d_l$ & $e^{il\pi}$ & $e^{i(l+1)\pi}$
      & $e^{il\pi}$ & $e^{i(l+1)\pi}$ \\
\hline\hline
\end{tabular}
\caption{Representations of the $C_2$ rotation at HSPs $\bd{\Gamma, M, X}$ and $\bd{Y}$ for band representations which are induced form the irreducible representations labelled by $l$ of the site symmetry groups for the maximal Wyckoff positions $a, b, c, d$ in $C_2$ symmetric lattices.$l=0,1$ for spinless TCIs and $l=\frac{1}{2},{3}{2}$ for spin-$\frac{1}{2}$ TCIs}
\label{Tab:Wy_C2}
\end{table}
where the angular momentum $l$ labels different irreducible representations of a site symmetry group. From Table~\ref{Tab:Wy_C2}, we count the number of the $p$-th ($p=1,2$) eigenvalue of $\{C_{2}|{0,0}\}$ at the HSPs $\bd{\Pi}^{(2)}$ from the irreducible representation $l$ of the site symmetry group of a maximal Wyckoff position $\alpha$ ($\alpha=a,b,c,d$). We denote this number as $l_{p\alpha}^{\bd{\Pi}^{(2)}}$. For example, we have $0_{1a}^{\bd{\Gamma}^{(2)}}=1$ and $0_{2a}^{\bd{\Gamma}^{(2)}}=0$ because each $a_{0}$ contributes only one $+1$ at the $\mathbf{\Gamma}$.
Note that in the $C_{2}$ symmetric TCIs, the site symmetry groups of Wyckoff positions a, b, c, d are all isomorphic to the point group $C_{2}$ which only has two (one-dimensional) irreducible representations. Thus, the angular momentum $l$ only has two values: $l=0,1$ for spinless TCIs and $l=\frac{1}{2},\frac{3}{2}$ for the spin-$\frac{1}{2}$ TCIs. For a TCI with the configuration of Wannier orbitals $\{n_{\alpha}^{l}\}$, at each HSP, we have $\sum_{\alpha,l} n_\alpha^{l} l_{p\alpha}^{\Pi^{(2)}}=\#\Pi_p^{(2)}$, where $p=1,2$. However, $4$ out of these $8$ equations are linearly independent because of the constraint $\sum_{\alpha,l}M_{\alpha} n_\alpha^l=\nu=\sum_p\#\Pi_p^{(n)}$, where $M_{\alpha}$ is the multiplicity of the Wyckoff position $\alpha$ and $\nu$ is the number of occupied states. With the constraint itself, we have $5$ linearly independent equations with $8$ unknowns $(n_{a}^{0},n_{b}^{0},n_{c}^{0},n_{d}^{0},n_{a}^{1},n_{b}^{1},n_{c}^{1},n_{d}^{1})$. Since the charge depends on the total number of Wannier orbitals at a Wykcoff position, we replace $n_{\alpha}^{0}$ with $n_{\alpha}=\sum_{l}n_{\alpha}^{l}$ for $\alpha=a,b,c,d$. For spinless TCIs these equations are
\begin{equation}
\label{eq:lineareqsetc2}
\begin{array}{l}
n_{a}+n_{b}+n_{c}+n_{d}=\nu,
\\
n_{a}^{0}+n_{b}^{0}+n_{c}^{0}+n_{d}^{0}=\#\Gamma^{(2)}_{1},
\\
n_{a}^{0}+n_{b}-n_{b}^{0}+n_{c}-n_{c}^{0}+n_{d}^{0}=\# X^{(2)}_{1},
\\
n_{a}^{0}+n_{b}-n_{b}^{0}+n_{c}^{0}+n_{d}-n_{d}^{0}=\# Y^{(2)}_{1},
\\
n_{a}^{0}+n_{b}^{0}+n_{c}-n_{c}^{0}+n_{d}-n_{d}^{0}=\# M^{(2)}_{1}.
\end{array}
\end{equation}
For spin-$\frac{1}{2}$ TCIs, we simply replace the integer superscript $l=0,1,\cdots$ by the half integer superscript $l=\frac{1}{2},\frac{3}{2},\cdots$. In the following sections for other $C_n$ symmetries, we only list the equations for spinless TCIs and the equations for spin-$\frac{1}{2}$ TCIs can be written in the same way as in here. Since we have more unknowns than equations, there is no unique solution. Leaving $n_{b}^{0}$, $n_{c}^{0}$ and $n_{d}^{0}$ to be unsolved, we get
\begin{equation}
\label{eq:wannierc2}
\begin{array}{l}
n_{b}=\frac{1}{2}\left(\left[X^{(2)}_{1}\right]+\left[ Y^{(2)}_{1}\right]-\left[M^{(2)}_{1}\right]+4n_{b}^{0}\right),
\\
\\
n_{c}=\frac{1}{2}\left(\left[X^{(2)}_{1}\right]-\left[ Y^{(2)}_{1}\right]+\left[M^{(2)}_{1}\right]+4n_{c}^{0}\right),
\\
\\
n_{d}=\frac{1}{2}\left(-\left[X^{(2)}_{1}\right]+\left[ Y^{(2)}_{1}\right]+\left[M^{(2)}_{1}\right]+4n_{d}^{0}\right),
\end{array}
\end{equation}
where, for obstructed atomic insulators or fragile TCIs, $n_b^0, n_c^0$ and $n_d^0$ can be arbitrary (possibly negative) integers. Thus, the undetermined parts of $n_b$, $n_c$ and $n_d$ are all multiples of $2$, which do \emph{not} affect the fractional portions of the bulk polarization and disclination charge according to Eq.~\eqref{eq:discharge} and Eq.~\eqref{eq:indexpolar}. Substituting Eq.~\eqref{eq:wannierc2} into Eq.~\eqref{eq:indexpolar} and Eq.~\eqref{eq:discharge}, for the bulk polarization we have
\begin{equation}
\label{eq:polarc2}
\begin{aligned}
\bold{P}^{(2)}&=\frac{1}{2}\left[(n_{b}+n_{c})\mathbf{e}_{1}+(n_{b}+n_{d})\mathbf{e}_{2}\right] \ \rm{mod} \ 1
\\
&=\frac{1}{2}\left(\left[X_{1}^{(2)}\right]\mathbf{e}_{1}+\left[Y_{1}^{(2)}\right]\mathbf{e}_{2}\right) \ \rm{mod} \ 1,
\end{aligned}
\end{equation}
as also found in Refs \onlinecite{benalcazar2019quantization,fang2012bulk}. For the disclination charge we have,
\begin{equation}
\label{eq:dischargec2}
\begin{aligned}
 \Delta&Q^{(2)}_{f}=\left[\frac{\Omega}{2\pi}(n_{b}+n_{c}+n_{d})+\mathbf{T}^{(2)}\cdot \mathbf{P}^{(2)}\right]\ \rm{mod} \ 1 
 \\
 &=\frac{\Omega}{2\pi}(\frac{1}{2}\left[X_{1}^{(2)}\right]+\frac{1}{2}\left[Y_{1}^{(2)}\right]+\frac{1}{2}\left[M_{1}^{(2)}\right])
 \\
 &+\mathbf{T}^{(2)}\cdot \mathbf{P}^{(2)}\ \rm{mod} \ 1.
\end{aligned}
\end{equation}
where ${\bf T}^{(2)}=a_1{\bf d}_1-a_2{\bf d}_2$, such that ${\bf d}_i\cdot {\bf e}_j=\delta_{i,j}$ for $i,j=1,2$ and $\mathbf{a}=a_{1}\mathbf{e}_{1}+a_{2}\mathbf{e}_{2}$ is the translation holonomy.

\subsection{Threefold symmetry}
\label{subsec:threefold}
For $C_{3}$ symmetric lattices, we choose the basis vectors to be $\mathbf{e}_{1}=(1,0)$ and $\mathbf{e}_{2}=(\frac{1}{2}, \frac{\sqrt{3}}{2})$. There are three maximal Wyckoff positions, which are labelled by a, b, and c as shown in Fig.~\ref{fig:WyckoffPos} in the main text.  They all have multiplicity $1$. Therefore, we can ignore the indices $\mu$ and $\nu$ in Eq.~\eqref{eq:bandrep}. The site symmetry group at each Wyckoff position is isomorphic to the $C_{3}$ point group which has three (one dimensional) irreducible representations. Therefore, we can also ignore the indices $i$, $j$. From Eq.\eqref{eq:bandrep}, we find
\begin{equation}
\label{eq:bandrepc3}
\begin{array}{l}
\rho_{G}^{\bold{k}}(\{C_{3}|(0,0)\})=\rho(\{C_{3}|(0,0)\}),
\\
\\
\rho_{G}^{\bold{k}}(\{C_{3}|(0,0)\})=e^{-i(R_{3}\bold{k})\cdot(-1,0)}\rho(\{C_{3}|(1,0)\}),
\\
\\
\begin{aligned}
\rho_{G}^{\bold{k}}(\{C_{3}|(0,0)\})=e&^{-i(R_{3}\bold{k})\cdot(-\frac{1}{2},\frac{\sqrt{3}}{2})}
\\
&\times\rho(\{C_{3}|(\frac{1}{2},-\frac{\sqrt{3}}{2})\}),
\end{aligned}
\end{array}
\end{equation}
for band representations induced from the irreducible representations of $\mathbf{G}_{a}$, $\mathbf{G}_{b},$ and $\mathbf{G}_{c}$, respectively. We then use Eq.~\eqref{eq:bandrepc3} to read off the eigenvalues that labels the representations of $\{C_{3}|(0,0)\}$ at all HSPs. The results are listed in Table~\ref{Tab:Wy_C3}. 
\begin{table}[h!]
\centering
\begin{tabular}{p{0.3\columnwidth}|p{0.2\columnwidth}p{0.2\columnwidth}p{0.2\columnwidth}}
\hline\hline
Wyckoff positions & $\bd{\Gamma}$ & $\bd{K}$ & $\bd{K'}$\\\hline
$a_l$   & $e^{\frac{i 2\pi l}{3}}$ & $e^{\frac{i 2\pi l}{3}}$ 
                & $e^{\frac{i 2\pi l}{3}}$\\
$b_l$ & $e^{\frac{i2l\pi}{3}}$ & $e^{\frac{i2(l+1)\pi}{3}}$ & $e^{\frac{i2(l-1)\pi}{3}}$\\
$c_l$ & $e^{\frac{i2l\pi}{3}}$ & $e^{\frac{i2(l-1)\pi}{3}}$ & $e^{\frac{i2(l+1)\pi}{3}}$\\
\hline\hline
\end{tabular}
\caption{Representations of the $C_3$ rotation at HSPs $\bd{\Gamma,K}$ and $\bd{K'}$ for band representations which are induced form the irreducible representations of the site symmetry group labelled by $l$ for the maximal Wyckoff positions $a,b,c$ in $C_3$ symmetric lattices. We have $l=0,1,2$ for the spinless TCIs and $l=\frac{1}{2},\frac{3}{2},\frac{5}{2}$ for the spin-$\frac{1}{2}$ TCIs.}
\label{Tab:Wy_C3}
\end{table}
Just as in the case of $C_2$ symmetric TCIs, we have $7$ linearly independent equations ($2$ from each HSP and $1$ from the constraint of the number of occupied states) for $9$ unknowns $(n_a,n_b,n_c,n_a^{0},n_b^{0},n_c^{0},n_a^{1},n_b^{1},n_c^{1})$. Reading the $l_{p\alpha}^{{\bf \Pi}^{(3)}}$ from Table~\ref{Tab:Wy_C3}, these equations are
\begin{equation}
\label{eq:lineareqsetc3}
\begin{array}{l}
n_{a}+n_{b}+n_{c}=\nu,
\\
n_{a}^{0}+n_{b}^{0}+n_{c}^{0}=\# \Gamma^{(3)}_{1},
\\
n_{a}^{1}+n_{b}^{1}+n_{c}^{1}=\# \Gamma^{(3)}_{2},
\\
n_{a}^{0}+n_{c}^{1}+n_{b}-n_{b}^{0}-n_{b}^{1}=\# K^{(3)}_{1},
\\
n_{a}^{1} + n_{c} - n_{c}^{0} - n_{c}^{1}+ n_{b}^{0} = \# K^{(3)}_{2},
\\
n_{a}^{0}+n_{c} - n_{c}^{0} - n_{c}^{1}+n_{b}^{1}=\# K^{\prime(3)}_{1},
\\
n_{a}^{1}+n_{c}^{0}+n_{b}-n_{b}^{0}-n_{b}^{1}=\# K^{\prime(3)}_{2}.
\end{array}
\end{equation}
Leaving $n_{b}^{1}$ and $n_{c}^{1}$ unsolved, we have 
\begin{equation}
\label{eq:wannierc3}
\begin{array}{l}
n_{b}=\left[K^{(3)}_{1}\right]+\left[K^{(3)}_{2}\right]-\left[K^{\prime(3)}_{1}\right]+3n_{b}^{1},
\\
\\
n_{c}=\left[K^{\prime (3)}_{1}\right]+\left[K^{\prime(3)}_{2}\right]-\left[K^{(3)}_{1}\right]+3n_{c}^{1}.
\end{array}
\end{equation}
Given that $n_{b}^1$ and $n_c^1$ are integers, the undetermined parts in $n_{b}$ and $n_{c}$ do \emph{not} affect the fractional portions of the disclination charge. By substituting Eq.\eqref{eq:wannierc3} into Eq.~\eqref{eq:indexpolar} and Eq.~\eqref{eq:discharge} in the main text, we have
\begin{equation}
\label{eq:dischargec3}
\begin{aligned}
 \Delta Q^{(3)}_{f}&=\left[\frac{\Omega}{2\pi}(n_{b}+n_{c})+\mathbf{T}^{(3)}\cdot\mathbf{P}^{(3)}\right] \rm{mod} \ 1
 \\
 &=\frac{\Omega}{2\pi}(\left[K_{2}^{(3)}\right]+\left[K^{\prime(3)}_{2}\right])+\mathbf{T}^{(3)}\cdot\mathbf{P}^{(3)}  \ \rm{mod} \ 1,
\end{aligned}
\end{equation}
where
\begin{equation}
\label{eq:polarc3}
\begin{aligned}
\bold{P}^{(3)}=&\frac{1}{3}\left[(n_{b}-n_{c})\mathbf{e}_{1}+(n_{b}-n_{c})\mathbf{e}_{2}\right] \ \rm{mod} \ 1
\\
=&\frac{1}{3}(\left[K_{2}^{(3)}\right]-\left[K^{\prime(3)}_{2}\right]
\\
&+2\left[K_{1}^{(3)}\right]-2\left[K^{\prime(3)}_{1}\right])(\bd{e}_1+\bd{e}_2) \ \rm{mod} \ 1.
\end{aligned}
\end{equation}
is the polarization in $C_{3}$ symmetric TCIs.
\subsection{Fourfold symmetry}
\label{subsec:fourfold}
For $C_{4}$ symmetric lattices, we choose the basis vectors to be $\mathbf{e}_{1}=(1,0)$ and $\mathbf{e}_{2}=(0,1)$. As shown in Fig.~\ref{fig:WyckoffPos}, there are two maximal Wyckoff positions, $a$ and $b$, both with multiplicity 1, and one maximal Wyckoff position $c$ with multiplicity 2. Thus, we cannot ignore the indices $\mu$ and $\nu$ in Eq.~\eqref{eq:bandrep} when studying the band representations induced from the irreducible representations of $\mathbf{G}_{c}$. Since $\mathbf{\Gamma}^{(4)}$ and $\mathbf{M}^{(4)}$ are invariant under $C_{4}$ rotation and $\mathbf{X}^{(2)}$ is invariant under $C_{2}$ rotation, we need to consider the induced band representations for both $C_2$ and $C_4$ rotation operators, $\rho_{\mathbf{G}}^{\mathbf{k}}(\{C_{2}|(0,0)\})$ and $\rho_{\mathbf{G}}^{\mathbf{k}}(\{C_{4}|(0,0)\})$. For representations induced from the irreducible representations of $\mathbf{G}_{a}$ and $\mathbf{G}_{b}$ (both isomorphic to the $C_{4}$ point group which has four (one-dimensional) irreducible representations), we have
\begin{align}
\label{eq:bandrepc4_1}
\rho_{G}^{\bold{k}}(\{C_{4}|(0,0)\})&=\rho(\{C_{4}|(0,0)\}),
\nonumber\\
\rho_{G}^{\bold{k}}(\{C_{2}|(0,0)\})&=\rho(\{C_{2}|(0,0)\}),
\nonumber\\
\rho_{G}^{\bold{k}}(\{C_{4}|(0,0)\})&=e^{-i(R_{4}\bold{k})\cdot(-1,0)}\rho(\{C_{4}|(1,0)\}),
\nonumber\\
\rho_{G}^{\bold{k}}(\{C_{2}|(0,0)\})&=e^{-i(R_{2}\bold{k})\cdot(-1,-1)}\rho(\{C_{2}|(1,1)\}).
\end{align}
For representations induced from
the irreducible representations of $\mathbf{G}_{c}=\{\{E|(0,0)\},\{C_{2}|(1,0)\}\}$, we have
\begin{align}
\label{eq:bandrepc4_2}
\rho_{G}^{\bold{k}}(\{C_{4}|(0,0)\})_{2,1}&=\rho(\{E|(0,0)\}),
\nonumber\\
\rho_{G}^{\bold{k}}(\{C_{4}|(0,0)\})_{1,2}&=e^{-i(R_{4}\bold{k})\cdot(-1,0)}\rho(\{C_{2}| (1,0)\}),
\nonumber\\
\rho_{G}^{\bold{k}}(\{C_{2}|(0,0)\})_{1,1}&=e^{-i(R_{2}\bold{k})\cdot(-1,0)}\rho(\{C_{2}|{1,0}\}),
\nonumber\\
\rho_{G}^{\bold{k}}(\{C_{2}|(0,0)\})_{2,2}&=e^{-i(R_{2}\bold{k})\cdot(0,-1)}\rho(\{C_{2}|{1,0}\}).
\end{align}
The subscripts in $[\rho_G^{\bf k}]_{i,j}$ are due to the nontrivial multiplicity of Wyckoff position $c$. From Eq.~\eqref{eq:bandrepc4_1} and Eq.~\eqref{eq:bandrepc4_2}, we can read off the eigenvalues that labels the representations of $\{C_{4}|(0,0)\}$ at $\mathbf{\Gamma}$ and $\mathbf{M}$, and $\{C_{2}|(0,0)\}$ at $\mathbf{X}$ and $\mathbf{Y}$ . The results are listed in Table~\ref{Tab:Wy_C4}.
\begin{table}[!t]
\centering
\begin{tabular}{p{0.3\columnwidth}|p{0.15\columnwidth}p{0.15\columnwidth}p{0.15\columnwidth}p{0.15\columnwidth}}
\hline\hline
Wyckoff positions& $\bd{\Gamma}$ & $\bd{M}$ & $\bd{X}$ & $\bd{Y}$\\
\hline
$a_l$ & $e^{\frac{i\pi l}{2}}$ & $e^{\frac{i\pi l}{2}}$
      & $e^{il\pi}$ & $e^{il\pi}$ \\
$b_l$ & $e^{\frac{i\pi l}{2}}$ & $-e^{\frac{i\pi l}{2}}$
	  & $e^{i(l+1)\pi}$ & $e^{i(l+1)\pi}$ \\
\multirow{2}{*}{$c_l$} 
     & $e^{\frac{i\pi }{2}(l+2)}$ & $e^{\frac{i\pi }{2}(l+1)}$ &$1$ or $i$ & $1$ or $i$\\
     & $ e^{\frac{i\pi l}{2}} $ & $e^{\frac{i\pi }{2}(l+3)}$ & $-1$ or $-i$ &$-1$ or $-i$\\
\hline\hline
\end{tabular}
\caption{Representations of $C_4$ ($C_{2}$) rotation at HSPs $\bd{\Gamma}$ and $\bd{M}$ ($\bd{X}$ and $\bd{Y}$) for band representations which are induced from the irreducible representations labelled by $l$ of site symmetry groups for the maximal Wyckoff positions $a$, $b$ and $c$ in $C_4$ symmetric lattices. The ``1 or $i$" means for the spinless case it is 1 and for the spin-$\frac{1}{2}$ case it is $i$. We have $l=0,1,\cdots, m_{\alpha}-1$ for the spinless case and $l=\frac{1}{2},\frac{3}{2},\cdots,m_{\alpha}-\frac{1}{2}$ for the spin-$\frac{1}{2}$ case. Here, we have $m_{\alpha}=4$ for $\alpha=a,b$ and $m_{c}=2$.}
\label{Tab:Wy_C4}
\end{table}
Then, with $l_{p\alpha}^{\bd{\Pi}^{(4)}}$ and $l_{p\alpha}^{\bd{\Pi}^{(2)}}$ counted from Table~\ref{Tab:Wy_C4} and the number of occupied bands $\nu$, we can have $8$ linearly independent equations for $\{n_{\alpha}^{l}\}$. By substituting $n_{\alpha}=\sum_{l}n_{\alpha}^{l}$ where $\alpha=a,b,c,$ into the $8$ linearly independent equations, we find
\begin{equation}
\label{eq:lineareqsetc4}
\begin{array}{l}
n_{a}+n_{b}+2n_{c}=\nu,
\\
n_{a}^{0}+n_{b}^{0}+n_{c}^{0}=\#\Gamma^{(4)}_{1},
\\
n_{a}^{1}+n_{b}^{1}+ (n_c -n_{c}^{0})=\#\Gamma^{(4)}_{2},
\\
n_{a}^{2}+n_{b}^{2}+n_{c}^{0}=\#\Gamma^{(4)}_{3},
\\
n_{a}^{0}+n_{b}^{2}+(n_{c} - n_{c}^{0})=\# M^{(4)}_{1},
\\
n_{a}^{1}+n_{b}-n_{b}^{0}-n_{b}^{1}-n_{b}^{2}+n_{c}^{0}=\# M^{(4)}_{2},
\\
n_{a}^{2}+n_{b}^{0}+(n_{c} -n_{c}^{0})=\# M^{(4)}_{3},
\\
n_{a}^{0}+n_{a}^{2}+n_{b}-n_{b}^{0}-n_{b}^{2}+n_{c}=\# X^{(2)}_{1}.
\end{array}
\end{equation}
Choosing $n_{c}^{0}$ and $n_{b}^{2}$ to be unsolved, we have
\begin{equation}
\label{eq:wannierc4}
\begin{array}{l}
n_{b}=\frac{1}{2}\left(2\left[X_{1}^{(2)}\right]+\left[M_{3}^{(4)}\right]-3\left[M_{1}^{(4)}\right]+8n_{b}^{2}\right),
\\
\\
n_{c}=\frac{1}{2}\left(\left[M_{1}^{(4)}\right]+\left[M_{3}^{(4)}\right]+4n_{c}^{0}\right).
\end{array}
\end{equation}
As before, given that $n_b^2$ and $n_c^0$ are integers, the undetermined parts in $n_b$ and $n_c$ do \emph{not} affect the fractional portions of the disclination charge. By substituting Eq.~\eqref{eq:wannierc4} into Eq.~\eqref{eq:indexpolar}, and Eq.~\eqref{eq:discharge} in the main text, we finally have
\begin{equation} 
\begin{aligned}
\Delta Q^{(4)}_{f}=&\left[\frac{\Omega}{2\pi}(n_{b}+2n_{c})+\mathbf{T}^{(4)}\cdot \mathbf{P}^{(4)}\right]\ \rm{mod} \ 1 
\\
=&\frac{\Omega}{2\pi}(\left[X_{1}^{(2)}\right]+\frac{3}{2}\left[M_{3}^{(4)}\right]-\frac{1}{2}\left[M_{1}^{(4)}\right])
\\
&+\mathbf{T}^{(4)}\cdot \mathbf{P}^{(4)}\ \rm{mod} \ 1,
\end{aligned}
\end{equation}
where
\begin{equation}
\begin{aligned}
\mathbf{P}^{(4)}&=\frac{1}{2}\left[(n_{b}+2n_{c})\mathbf{e}_{1}+(n_{b}+2n_{c})\mathbf{e}_{2}\right] \ \rm{mod} \ 1
\\
&=\frac{1}{2}(\left[X_{1}^{(2)}\right]+\left[M_{3}^{(4)}\right]-\left[M_{1}^{(4)}\right])(\mathbf{e}_{1}+\mathbf{e}_{2}) \ \rm{mod} \ 1.
\end{aligned}
\end{equation}
is the polarization in $C_{4}$ symmetric TCIs.

\subsection{Sixfold symmetry}
For $C_{6}$ symmetric lattices, we choose $\mathbf{e}_{1}=(1,0)$, $\mathbf{e}_{2}=(\frac{1}{2}, \frac{\sqrt{3}}{2})$ to be the basis vectors of the lattice. As shown in Fig.~\ref{fig:WyckoffPos} in the main text, we have three maximal Wyckoff positions, $a$ with multiplicity $1$, $b$ with multiplicity $2,$ and $c$ with multiplicity $3$. We need to find the induced band representation $\rho_{\mathbf{G}}^{\mathbf{k}}$ of $\{C_{6}(0,0)\}$, $\{C_{3}(0,0)\},$ and $\{C_{2}(0,0)\}$ from the irreducible representations $\rho$ of the site symmetry group $G_\alpha, \alpha = a, b, c$. The induced band representation from the irreducible representation $\rho$ of $\mathbf{G}_{a}$ ($C_{6}$ point group) is simply $\rho_{\mathbf{G}}^{\mathbf{k}}(h)=\rho(h)$ for $\forall h\in \mathbf{g}$. Using Eq.~\eqref{eq:bandrep}, the induced representations from the irreducible representations of $\mathbf{G}_{b}=\{\{E|(0,0)\},\{C_{3}|(1,0)\}, \{C_{3}^{2}|(\frac{1}{2},\frac{\sqrt{3}}{2})\}\}$ are
\begin{equation}
\label{eq:bandrepc6_1}
\begin{array}{l}
\rho_{\mathbf{G}}^{\mathbf{k}}(\{C_{6}|(0,0)\})_{2,1}=\rho(\{E|(0,0)\}),
\\
\rho_{\mathbf{G}}^{\mathbf{k}}(\{C_{6}|(0,0)\})_{1,2}=e^{-i(R_{6}\mathbf{k})\cdot (-1,0)}\rho(\{C_{3}|(1,0)\}),
\\
\rho_{\mathbf{G}}^{\mathbf{k}}(\{C_{3}|(0,0)\})_{1,1}=e^{-i(R_{3}\mathbf{k})\cdot (-1,0)}\rho(\{C_{3}|(1,0)\}),
\\
\rho_{\mathbf{G}}^{\mathbf{k}}(\{C_{3}|(0,0)\})_{2,2}=e^{-i(R_{3}\mathbf{k})\cdot (-\frac{1}{2},-\frac{\sqrt{3}}{2})}\rho(\{C_{3}|(1,0)\}),
\\
\rho_{\mathbf{G}}^{\mathbf{k}}(\{C_{2}|(0,0)\})_{2,1}=e^{-i(R_{2}\mathbf{k})\cdot (-\frac{1}{2},-\frac{\sqrt{3}}{2})}\rho(\{C_{3}|(1,0)\}),
\\
\rho_{\mathbf{G}}^{\mathbf{k}}(\{C_{2}|(0,0)\})_{1,2}=e^{-i(R_{2}\mathbf{k})\cdot (-\frac{1}{2},-\frac{\sqrt{3}}{2})}\rho(\{C_{3}^{2}|(\frac{1}{2},
\frac{\sqrt{3}}{2})\}).
\end{array}
\end{equation}
Similarly, the induced band representations from the irreducible representations of $\mathbf{G}_{c}=\{\{E|(0,0)\},\{C_{2}|(1,0)\}\}$ are
\begin{equation}
\label{eq:bandrepc6_2}
\begin{array}{l}
\rho_{\mathbf{G}}^{\mathbf{k}}(\{C_{6}|(0,0)\})_{3,1}=e^{-i(R_{6}\mathbf{k})\cdot (\frac{1}{2},\frac{\sqrt{3}}{2})}\rho(\{C_{2}|(1,0)\}),
\\
\rho_{\mathbf{G}}^{\mathbf{k}}(\{C_{6}|(0,0)\})_{1,2}=e^{-i(R_{6}\mathbf{k})\cdot (-1,0)}\rho(\{C_{2}|(1,0)\}),
\\
\rho_{\mathbf{G}}^{\mathbf{k}}(\{C_{6}|(0,0)\})_{2,3}=e^{-i(R_{6}\mathbf{k})\cdot (\frac{1}{2},-\frac{\sqrt{3}}{2})}\rho(\{C_{2}|(1,0)\}),
\\
\rho_{\mathbf{G}}^{\mathbf{k}}(\{C_{3}|(0,0)\})_{2,1}=\rho(\{E|(1,0)\}),
\\
\rho_{\mathbf{G}}^{\mathbf{k}}(\{C_{3}|(0,0)\})_{3,2}=\rho(\{E|(1,0)\}),
\\
\rho_{\mathbf{G}}^{\mathbf{k}}(\{C_{3}|(0,0)\})_{1,3}=\rho(\{E|(1,0)\}),
\\
\rho_{\mathbf{G}}^{\mathbf{k}}(\{C_{2}|(0,0)\})_{1,1}=e^{-i(R_{2}\mathbf{k})\cdot (-1,0)}\rho(\{C_{2}|(1,
0)\}),
\\
\rho_{\mathbf{G}}^{\mathbf{k}}(\{C_{2}|(0,0)\})_{2,2}=e^{-i(R_{2}\mathbf{k})\cdot (\frac{1}{2},-\frac{\sqrt{3}}{2})}\rho(\{C_{2}|(1,
0)\}),
\\
\rho_{\mathbf{G}}^{\mathbf{k}}(\{C_{2}|(0,0)\})_{3,3}=e^{-i(R_{2}\mathbf{k})\cdot (\frac{1}{2},\frac{\sqrt{3}}{2})}\rho(\{C_{2}|(1,
0)\}).
\end{array}
\end{equation}
From Eq.~\eqref{eq:bandrepc6_1} and Eq.~\eqref{eq:bandrepc6_2}, we can read off the eigenvalues that labels the representations of $\{C_{6}|(0,0)\}$ at $\mathbf{\Gamma}$, $\{C_{3}|(0,0)\}$ at $\mathbf{K}$ and $\{C_{2}|(0,0)\}$ at $\mathbf{M}$. The results are listed in Table~\ref{Tab:Wy_C6}.
With $l^{\bd{\Pi}^{(6)}}_{p\alpha}$, $l^{\bd{\Pi}^{(3)}}_{p\alpha},$ and $l_{p\alpha}^{\bd{\Pi}^{(2)}}$ counted from Table~\ref{Tab:Wy_C6}, we have $9$ linearly independent equations ($5$ at the ${\bf \Gamma}$ point, $2$ at the ${\bf K}$ point, $1$ at the ${\bf M}$ point, and $1$ from the constraints of the number of occupied states $\nu$) for $11$ unknowns $(n_a^l,l=0,\ldots5,n_b^l,l=0,\ldots,2,n_c^l,l=0,1) $. By replacing $n_\alpha^{M_\alpha-1}$ with $n_{\alpha}=\sum_{l}n_{\alpha}^{l},l=0,\ldots M_\alpha -1$ for $\alpha=a,b,c$, we have
\begin{equation}
\label{eq:lineareqsetc6}
\begin{array}{l}
n_{a}+2n_{b}+3n_{c}=\nu,
\\
n_{a}^{0}+n_{b}^{0}+n_{c}^{0}=\#\Gamma^{(6)}_{1},
\\
n_{a}^{1}+n_{b}^{1}+(n_{c} - n_{c}^{0})=\#\Gamma^{(6)}_{2},
\\
n_{a}^{2}+(n_{b}-n_{b}^{0}-n_{b}^{1})+n_{c}^{0}=\#\Gamma^{(6)}_{3},
\\
n_{a}^{3}+n_{b}^{0}+(n_{c} -n_{c}^{0})=\#\Gamma^{(6)}_{4},
\\
n_{a}^{4}+n_{b}^{1}+n_{c}^{0}=\#\Gamma^{(6)}_{5},
\\
n_{a}^{0}+n_{a}^{3}+(n_{b}-n_{b}^{0}-n_{b}^{1})+n_{b}^{1}+n_{c}=\# K^{(3)}_{1},
\\
n_{a}^{1}+n_{a}^{4}+n_{b}^{0}+(n_{b}-n_{b}^{0}-n_{b}^{1})+n_{c}=\# K^{(3)}_{2},
\\
n_{a}^{0}+n_{a}^{2}+n_{a}^{4}+n_{b}+n_{c}^{0} + 2 (n_{c} - n_{c}^{0})=\# M^{(2)}_{1}.
\end{array}
\end{equation}
Here we choose $n_{c}^{0}$ and $n_{b}^{1}$ to be unsolved. Then $n_b$ and $n_c$ can be written as
\begin{equation}
\label{eq:wannierc6}
\begin{array}{l}
n_{b}=\left[K_{2}^{(3)}\right]+3n_{b}^{1},
\\
\\
n_{c}=\frac{1}{2}\left[M^{(2)}_{1}\right]+2n_{c}^{0}.
\end{array}
\end{equation}
As before, given $n_b^1$ and $n_c^0$ are integers, the undetermined parts in $n_b$ and $n_c$ do \emph{not} affect the fractional portions of the disclination charge. 
By substituting Eq.~\eqref{eq:wannierc6} into Eq.~\eqref{eq:discharge} in the main text, we have
\begin{equation}
\label{eq:dischargec6}
\Delta Q_{f}^{(6)}=\frac{\Omega}{2\pi}\left(2\left[K_{2}^{(3)}\right]+\frac{3}{2}\left[M^{(2)}_{1}\right]\right) \ \rm{mod} \ 1.
\end{equation}
\begin{table}[t!]
\centering
\begin{tabular}{p{0.3\columnwidth}|p{0.2\columnwidth}p{0.2\columnwidth}p{0.2\columnwidth}}
\hline\hline
Wyckoff positions & $\bd{\Gamma}$ & $\bd{K}$ & $\bd{M}$\\\hline
$a_l$   & $e^{\frac{i \pi l}{3}}$ & $e^{\frac{i 2\pi l}{3}}$ 
                & $e^{il\pi}$\\
\multirow{2}{*}{$b_l$} 
&$e^{\frac{i l \pi}{3}}$ & $e^{\frac{-i2\pi(l-1)}{3}}$ & $1$ or $i$\\
&$-e^{\frac{i l \pi}{3}}$ & $e^{\frac{-i2\pi(l+1)}{3}}$ & $-1$ or $-i$\\
\multirow{2}{*}{$c_l$}  
& $e^{i\frac{l\pi}{3}}$ & $1$ or $i$ & $e^{il\pi}$\\
& $-e^{i\frac{(l\pm1)\pi}{3}}$ & $e^{\pm\frac{i2\pi}{3}}$ & $e^{i(l\pm 1)\pi}$\\
\hline\hline
\end{tabular}
\caption{Eigenvalues for $C_6$ rotation operator at $\bd{\Gamma}$ points, $C_3$ rotation operator at $\bd{K}$ points and $C_2$ rotation operator at $\bd{M}$ points for bands which form the irreducible representations labelled by $l$ of site symmetry groups in $C_6$ symmetric lattices. The ``1 or $i$" means that for the spinless case it is 1 and for the spin-$\frac{1}{2}$ case it is $i$. We have $l=0,1,\cdots, m_{\alpha}-1$ for the spinless case and $l=\frac{1}{2},\frac{3}{2},\cdots,m_{\alpha}-\frac{1}{2}$ for the spin-$\frac{1}{2}$ case. Here, we have $m_{a}=6$, $m_{b}=3$ and $m_{c}=2$.}
\label{Tab:Wy_C6}
\end{table}
\section{Numerical results of fractional disclination charge for Chern insulators}
\label{sec: numchern}
\begin{figure}[t!]
\centering
\includegraphics[width=\columnwidth]{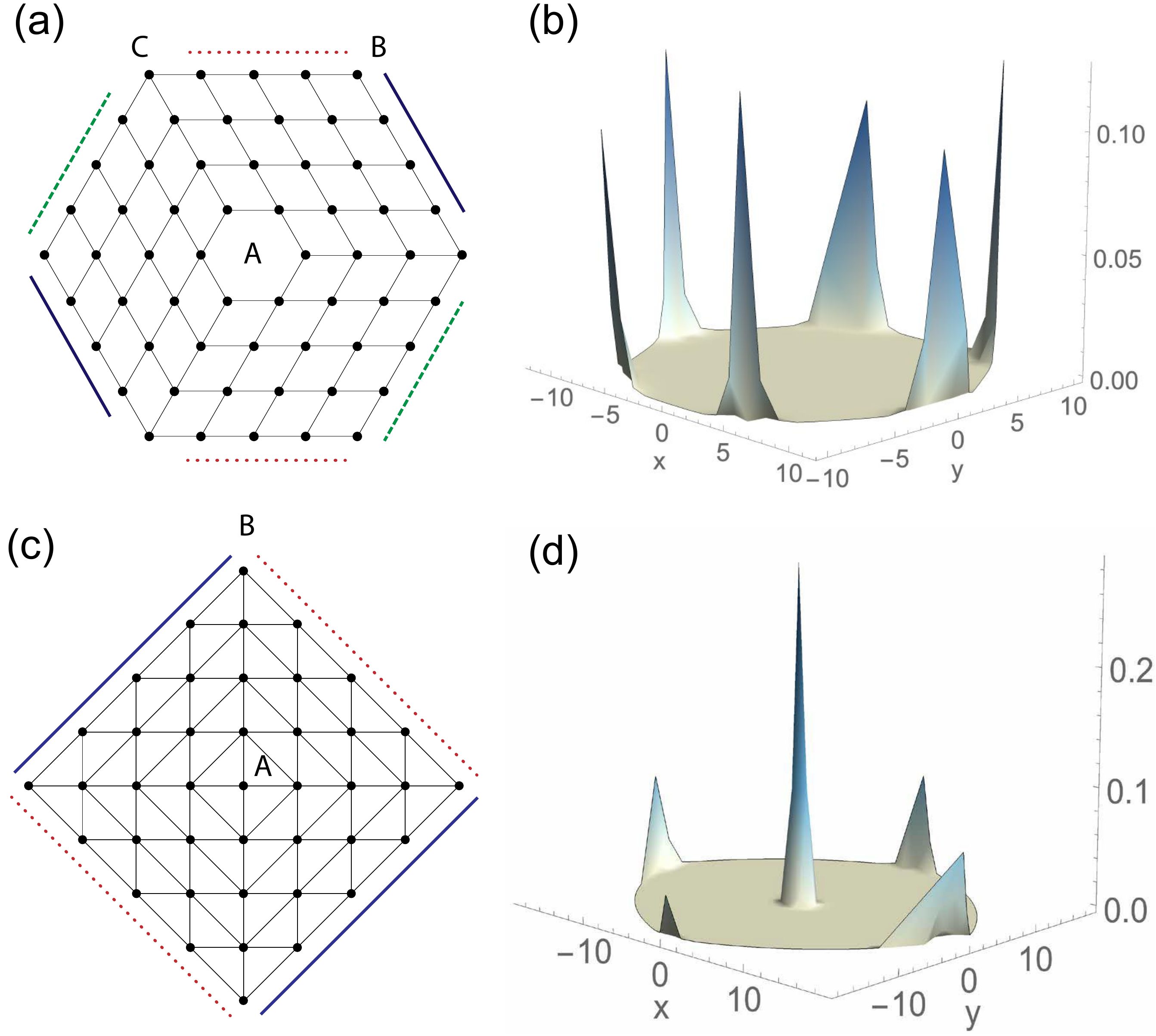}
\caption{(Color online) (a) Lattice configuration of the model in Eq.~\eqref{eq:chernc4}. Periodic boundary conditions are imposed by connecting the boundaries of the same color and line pattern with hopping terms as in the bulk. There are three disclinations in this lattice, which are centered around $A, B$ and $C$. Disclination $A$ has $n^{(4)}_{\Omega}=-1,[\bd{a}]^{(4)}=0$; $B$ has $n^{(4)}_{\Omega}=-1,[\bd{a}]^{(4)}=1$; and $C$ has $n^{(2)}_{\Omega}=1,[\bd{a}]^{(2)}=(1,0)$. (b) Charge density for Eq.~\eqref{eq:chernc4} on the lattice in (a) with $396$ unit cells and $m=1$. (c) Lattice configuration of the Haldane model in Eq.~\eqref{eq:chernc6}. Periodic boundary conditions are imposed by connecting the boundaries of the same color by hopping terms as in the bulk.
There are two disclinations in this lattice, which are centered around $A$ and $B$. Disclination $A$ has $n^{(3)}_{\Omega}=-1,[\bd{a}]^{(3)}=0$ and $B$ has $n^{(3)}_{\Omega}=1,[\bd{a}]^{(3)}=0$. (d) Charge density distribution for  Eq.~\eqref{eq:chernc6} on the lattice in (c) with $724$ unit cells, $t=1$, and $\lambda=0.2$.}
\label{fig:Chern_dis}
\end{figure}
For $C_{n}$ symmetric TCIs with non-zero Chern numbers, the (generalized) Wannier representation does not apply.
In this section, we numerically simulate $C_n$ symmetric Chern insulators with disclinations and calculate the bound fractional disclination charge. We use a two-band $C_{4}$ symmetric Chern-insulator model (naturally, this model is also $C_2$ symmetric) with  Hamiltonian
\begin{equation}
\label{eq:chernc4}
\begin{aligned}
H^{(4)}_{ch}&=\frac{1}{2}\sum_{x,y}\bigg(mc_{x,y}^{\dag}c_{x,y}+c_{x+1,y}^{\dag}c_{x,y}\otimes(i\sigma_
{x}+\sigma_{z})
\\
&+c_{x,y+1}^{\dag}c_{x,y}\otimes (i\sigma_{y}+\sigma_{z})+h.c.\bigg)
\end{aligned}
\end{equation}
and the two-band $C_{6}$ symmetric Haldane model \cite{haldane1988model} (naturally, this model is also $C_3$ symmetric):
\begin{equation}
\label{eq:chernc6}
H^{(6)}_{ch}=-t\sum_{\langle i,j\rangle}c_{i}^{\dag}c_{j}+\lambda\sum_{\langle\langle i,j\rangle\rangle}i\theta_{ij}c^{\dag}_{i}c_{j}，
\end{equation}
where $\langle i,j\rangle$ and $\langle\langle i,j\rangle\rangle$ refer to the nearest neighbors and the next nearest neighbors, respectively. The parameters $\theta_{i,j}$ are $+1 (-1)$ if the hopping from $j$ to $i$ is anti-clockwise (clockwise). When $0<m<2 \,\,(-2<m<0)$, $H^{(4)}_{ch}$ has Chern number $+1(-1),$ and when $\lambda>0 (\lambda<0)$, $H^{(6)}_{ch}$ has Chern number $+1(-1)$.
For $H_{ch}^{(4)}$ we use a periodic lattice with three disclinations centered at $A, B$ and $C$ as shown in Fig.~\ref{fig:Chern_dis}(a). The boundaries of the same color are connected by hopping terms. 

The disclination centered at $A$ has $\Omega=-\frac{\pi}{2},[\bd{a}]^{(4)}=0$; the disclination centered at $B$ has $\Omega=-\frac{\pi}{2},\left[\bd{a}\right]^{(4)}=1$; and the disclination centered at $C$ has $\Omega=\pi,\left[\bd{a}\right]^{(2)}=(1,0)$. 
For $H_{ch}^{(6)}$, we use the periodic lattice with two disclinations $A$ and $B$ as shown in Fig.~\ref{fig:Chern_dis}(c). The boundaries of the same color are connected by hopping terms. The disclination centered at $A$ has $\Omega=-\frac{2\pi}{3},[\bd{a}]^{(3)}=0$ and the disclination centered at $B$ has $\Omega=\frac{2\pi}{3},[\bd{a}]^{(3)}=0$.
Figs.~\ref{fig:Chern_dis} (b, d) show the fractional portions of charge density distributions when $ch=1$ in lattices shown in Fig.~\ref{fig:Chern_dis} (a, c), respectively. We observe fractional charge exponentially localized at the disclination cores. By integrating the charge density over a finite area around the disclination core, we verify that fractional disclination charge is quantized. We summarize all the numerical results of the quantized fractional disclination charge in Tables ~\ref{tab:c2c4ch} and ~\ref{tab:c3c6ch}.
\\
\begin{table}[H]
\centering
\renewcommand{\arraystretch}{1.5}
\begin{tabular}{p{0.08\columnwidth}|p{0.456\columnwidth}p{0.456\columnwidth}}
\hline\hline
\multirow{2}{*}{$C_2$} 
       &$\Omega=\pi$, $ch=1$ & $\Omega=\pi$, $ch=-1$ \\
     & $[{\bf a}]^{(2)}=(1,0)$ & $[{\bf a}]^{(2)}=(1,0)$\\
\hline
$Q_{dis}^{(2)}$ &  \ \ $\frac{1}{2}$  &  \ \ $0$ \\
\hline
\end{tabular}
\begin{tabular}{p{0.08\columnwidth}|p{0.22\columnwidth}p{0.22\columnwidth}p{0.22\columnwidth}p{0.22\columnwidth}}
\hline
\multirow{2}{*}{$C_4$} 
      & $\Omega =-\frac{\pi}{2}$ & $\Omega= -\frac{\pi}{2}$  
      & $\Omega=-\frac{\pi}{2}$    & $\Omega=-\frac{\pi}{2}$\\
      & $ch=1$ & $ch=1$  
      & $ch=-1$    & $ch=-1$\\
      & $[{\bf a}]^{(4)}=0$      & $[{\bf a}]^{(4)}=1$      & $[{\bf a}]^{(4)}=0$ 
      & $[{\bf a}]^{(4)}=1$\\
\hline
$Q_{dis}^{(4)}$ & \ \ 0 & \ \ $\frac{1}{2}$ & \ \ 0 & \ \ 0\\ 
\hline\hline
\end{tabular}
\caption{Fractional charge trapped at the core of different disclinations in $C_2$ and $C_4$ symmetric Chern insulators.}
\label{tab:c2c4ch}
\end{table}
\begin{table}[H]
\centering
\renewcommand{\arraystretch}{1.5}
\begin{tabular}{p{0.08\columnwidth}|p{0.22\columnwidth}p{0.22\columnwidth}p{0.22\columnwidth}p{0.22\columnwidth}}
\hline\hline
\multirow{2}{*}{$C_3$} 
       &$\Omega=-\frac{2\pi}{3}$ & $\Omega=\frac{2\pi}{3}$ & $\Omega=-\frac{2\pi}{3}$& $\Omega=\frac{2\pi}{3}$ \\
        & $ch=1$ & $ch=1$  
      & $ch=-1$    & $ch=-1$\\
     & $[{\bf a}]^{(3)}=0$ & $[{\bf a}]^{(3)}=1$& $[{\bf a}]^{(3)}=0$ & $[{\bf a}]^{(3)}=1$\\
\hline
$Q_{dis}^{(3)}$ &  \ \ $\frac{1}{2}$  &  \ \ $\frac{1}{2}$ &  \ \ $\frac{1}{2}$ &  \ \ $\frac{1}{2}$ \\
\hline
\end{tabular}
\begin{tabular}{p{0.08\columnwidth}|p{0.456\columnwidth}p{0.456\columnwidth}}
\hline
\multirow{2}{*}{$C_6$} 
      & $\Omega =-\frac{\pi}{3}$  & $\Omega=-\frac{\pi}{3}$   \\
      & $ch=1$   & $ch=-1$  \\
\hline
$Q_{dis}^{(6)}$ & \ \ $\frac{1}{4}$ & \ \ $\frac{1}{4}$   \\
\hline\hline
\end{tabular}
\caption{Fractional charge trapped at the core of different disclinations in $C_3$ and $C_6$ symmetric Chern insulators.}
\label{tab:c3c6ch}
\end{table}

\end{document}